\documentclass[aps, prd, reprint, longbibliography, nofootinbib,superscriptaddress,floatfix]{revtex4-2}

\pdfoutput=1

\usepackage{graphicx} 
\usepackage{xcolor}
\usepackage{babel}
\usepackage{amsthm,amsmath,amssymb,mathrsfs,amsfonts}
\usepackage{bbm}
\usepackage{slashed}
\usepackage{verbatim}
\usepackage[T1]{fontenc}
\usepackage[utf8]{inputenc}
\usepackage[colorlinks=true,linktocpage=true,linkcolor=blue,citecolor=blue]{hyperref}
\usepackage{mathbbol}
\usepackage[normalem]{ulem}
\usepackage{svg}
\usepackage{lipsum}
\usepackage{physics}
\usepackage{dcolumn}
\usepackage{bm}
\usepackage{tensor}
\usepackage{comment}
\usepackage{color,overpic,mathtools}
\usepackage{braket,setspace}
\usepackage{cancel}
\usepackage{float}
\usepackage{fancyhdr}
\fancypagestyle{plain}

\newcommand{\br}[1]{\left(#1\right)}

\newcommand{\p}{\pi}

\begin{document}


\title{Gravitational collapse in Quadratic Gravity}

\author{Ramiro Cayuso}
\affiliation{Perimeter Institute for Theoretical Physics, 31 Caroline Street North, Waterloo, ON N2L 2Y5, Canada}
\affiliation{Department of Physics and Astronomy, University of Waterloo, Waterloo, Ontario, N2L 3G1, Canada}

\begin{abstract}
This study explores the gravitational collapse of a massless scalar field within Quadratic Gravity treated as a dimension-four operator Effective Field Theory extension to General Relativity. The additional degrees of freedom associated with the higher derivatives in this theory are removed by an Order Reduction approach, where the truncated expansion nature of the theory is exploited. Through simulations, we find scenarios where solutions remain within the bounds of the Effective Field Theory while displaying significant deviations from General Relativity in the dynamics of curvature invariants during the collapse. Limitations of the approach taken, the Effective Field Theory approximation, and the appearance of instabilities are also discussed.
\end{abstract}
\maketitle

\section{Introduction}

Gravitational Wave (GW) Astronomy \cite{LIGOScientific:2021djp,LIGOScientific:2016aoc} has emerged as an extraordinary tool for probing the
nature of gravity via a channel and regimes that were inaccessible before
its time. By collecting and analyzing gravitational wave data from current 
and future detectors, we will be able to test General Relativity (GR) \cite{LIGOScientific:2021sio} with 
scrutiny limited only by the reach and precision of our detectors, as well as the quality 
of our predictions.
In the search for deviations from GR, the community has developed many alternative theories of gravity, for which substantial theoretical 
efforts have been placed into modeling and predictions. GW signals produced in compact binary mergers
are arguably the best source to peer into GR and possible modifications in the most dynamical and strong regime.
There are now several instances \cite{Cayuso:2023aht, Corman:2022xqg, AresteSalo:2022hua, Bezares:2021dma, Barausse:2012da, Elley:2022ept, Figueras:2021abd, Hirschmann:2017psw} where full nonlinear numerical simulations of compact binary coalescence (and the prediction of their respective GW emissions) have been achieved in modified gravity candidate theories. Understanding how modifications in the underlying theory change predictions is essential in pushing our searches for such deviations in the data.

Of the proposed theories which could be tested through the observation of gravitational waves, there is great interest in those that fall under
what is commonly called Effective Field Theory (EFT) extensions to GR \cite{Endlich:2017tqa, Ruhdorfer:2019qmk, Burgess:2003jk, Donoghue:1994dn}. These
theories are constructed by adding terms to the Einstein-Hilbert action formed from powers of curvature invariants that are adequately suppressed by powers of a given cut-off scale $\Lambda$. The scale $\Lambda$ is related to the mass of the heavy fields modifying the theory, which in the EFT description are integrated out. This method then describes a perturbative expansion consistent with the desired symmetries and assumptions, without introducing new light degrees of freedom. In recent years there have been several efforts \cite{Cano:2021myl, PhysRevD.102.044056,deRham:2021bll, EFTIR, Cayuso:2023aht} in the modeling of these theories, and constraining the relevant parameters, such as the scale $\Lambda$ at which modifications are introduced. These have mainly been focused on theories built using either six-dimensional 
or eight-dimensional operators, built from the contractions of three and four Riemann tensors. These are the leading and next-to-leading order operators in the absence of matter.

When matter is present, the leading order curvature operators in the EFT construction are dimension-four operators ($ R^2$, $R_{ab}R^{ab}$ and $R_{abcd}R^{abcd} $). In this context, neutron star (NS) binaries become one of the most relevant scenarios. Modifications to GR may not only affect the dynamics during the inspiral and merger phases but the behavior and signatures of the merger remnant could also be highly altered.
Given that these theories are constructed from powers of curvature invariants, it is natural that the effects of the modifications grow with the curvature, and small black holes (BHs) would give rise to the strongest effects. The merger of binary NSs \cite{LIGOScientific:2017vwq} presents an ideal scenario for the formation of some of the smallest astrophysical black holes, with masses of approximately $3M_\odot$. The post-merger dynamics of such an object could be one of the best windows to observe deviations from GR \cite{https://doi.org/10.48550/arxiv.2208.09488}. Exotic formation channels for smaller BHs could result in scenarios where such BHs interact with NSs in regimes of large spacetime curvature, where significant corrections could arise from these types of modifications to GR \footnote{See \cite{East:2019dxt} for a study of a NS being consumed by a much less massive BH residing inside the star}.

The theory built from these four-dimensional operators is commonly called Quadratic Gravity \cite{Stelle:1976gc}, and there has been recent work performing fully nonlinear numerical simulations in spherical symmetry and very recently in the BH binary merger scenario \cite{Held:2021pht, Held:2023aap}. However, these works have focused on the vacuum scenario, most specifically in the Ricci-flat case, which, from the perspective of EFT, solutions and dynamics should be indistinguishable from GR.

This work explores the dynamics of this dimension-four operator EFT extension to GR in the presence of matter, where modifications should arise. For simplicity, the considered system has spherical symmetry, and we evolve the collapse of a massless minimally coupled scalar field into a BH. There are several objectives to this work. First, we want to present an alternative approach to that presented in \cite{Held:2021pht, Held:2023aap}, as well as incorporate matter into the system to study gravitational collapse. The second one is to study how the modifying terms affect the dynamics of the system. And finally, to determine in what region of the parameter space the system stays within the EFT description, simulations are well-behaved, and when their predictions can be trusted. 

The paper is structured as follows: In section \ref{sfieldseqs}, the four-dimensional operator EFT, its action, and its corresponding field equations are presented. In section \ref{evolution_equations}, the evolution and constraint equations are presented, and the ``Order Reduction'' procedure is introduced to deal with the higher derivatives in such equations. Section \ref{target} contains detailed information about the target problem and setup, including the prescription for initial data, the numerical implementation, and relevant monitoring quantities. The main results of the paper are presented in Section \ref{results}. A brief discussion on the observed results and future outlook can be found in Section \ref{discussion}. The appendices contain additional information regarding the convergence test and constraint violations observed in the simulations. The following notation is adopted: The beginning of the Latin alphabet $(a, b, c, d,...)$ will be used to denote full spacetime indices, while the Latin letters $(i, j, k, l...)$ will be used to indicate spatial ones. The $(-, +, +, +)$ signature is used, and the speed of light is set to $c=1$.

\section{Leading order EFT, non vacuum equations}\label{sfieldseqs}
The leading order terms in an EFT extension to GR, which introduce no new light degrees of freedom and satisfy parity symmetry, are the ones built with the dimension-four operator curvature invariants $R^2$, $R_{a  b}R^{a b}$ and $R_{a b c d}R^{a b c d}$. Using the fact that the Gauss-Bonnet invariant is topological in four spacetime dimensions, one can exclude the Riemann-squared term from the effective action. The effective action can be written as:  
\begin{equation}
    S_\text{eff} = \frac{1}{16\p G}\int d^4x\, \sqrt{-g}\br{R -\frac{a_1}{\Lambda^{2}}R_{a b}R^{a b} + \frac{a_2}{\Lambda^{2}}R^2  + \cdots}\, ,
\end{equation}
where $a_{1}$ and $a_{2}$ are dimensionless coefficients and $\Lambda$ has units of inverse length and determines the cut-off of the EFT. Notice that in the vacuum case, since $R_{a b} = 0 + \mathcal{O}(1/\Lambda^2)$, then these terms would be pushed to higher orders of the perturbative scheme, and six-dimensional operators would dominate. This work includes matter in the form of a minimally coupled scalar field, so these terms are the leading order operators. 

Upon variation of this action, the following field equations are obtained,
\begin{align}
\begin{split}
    &R_{a b} -\frac{1}{2}g_{a b}R + \frac{1}{2}\epsilon_{1}R_{c d}R^{c d}g_{a b} +2\epsilon_{2}R_{a b}R -\frac{1}{2}\epsilon_{2}g_{a b}R^{2}\\
    &- 2\epsilon_{1}R^{c d}R_{acbd} +(\epsilon_{1} - 2\epsilon_{2})\nabla_{b}\nabla_{a}R    -\epsilon_{1}\nabla^{2}R_{ab} \\ 
    & - g_{a b}(\frac{1}{2}\epsilon_{1}-2\epsilon_{2})\nabla_{c}\nabla^{c}R = 8\pi T_{ab} , \label{Field_eqs_o}
\end{split}
\\&\nabla^{a}T_{a b} = 0 ,
\end{align}
where $\epsilon_{1} = a_1/\Lambda^2$, $\epsilon_{2} = a_2/\Lambda^2$ (which will occasionally be called couplings) and $T_{a b}$ is the usual energy-momentum tensor defined as,

\begin{equation}
   T_{a b} = \nabla_{a}\phi \nabla_{b}\phi - \frac{1}{2}g_{a b}\nabla_{c}\phi\nabla^{c}\phi. 
\end{equation}

For convenience equation \eqref{Field_eqs_o} will expressed as,

\begin{equation}
R_{a b} -\frac{1}{2}g_{a b}R  = 8\pi T_{ab} +  M_{a b},
\end{equation}
where now $M_{ab}$ encompasses all modifications to the equations. The $M_{a b}$ tensor contains up to 4th-order derivatives of the metric; this sort of modification makes the task of formulating the problem as well-posed  \cite{hadamard,Sarbach:2012pr} a challenging task, if not an impossible one with the standard techniques.

\section{Evolution equations and constraints}\label{evolution_equations}
Before addressing the issues raised at the end of the previous section, the equations will be first expressed in a formulation that, in the absence of correcting terms, renders the problem well-posed. To this end, the Generalized Harmonic formulation \cite{Foures-Bruhat:1952grw, Pretorius:2004jg, Lindblom:2005qh} that is written in terms of the usual 3+1 variables \cite{Brown:2011qg} is adopted. Under this formulation, the full set of evolution equations and constraints are expressed as, 

\begin{subequations}
\label{system_M}
\begin{align}
\partial_{\perp} \gamma_{i j}=&-2 \alpha K_{i j}, \label{subeq1}
\\
\begin{split}
\partial_{\perp} K_{i j}  = & \alpha\left[R^{(3)}_{i j}-2K_{i k} K_{j}^{k}-\widetilde{\pi} K_{i j}\right]-D_{i} D_{j} \alpha \\& -\alpha D_{(i} \mathcal{C}_{j)}-\kappa \alpha \gamma_{i j} \mathcal{C}_{T} / 2 \\ 
& -8 \pi G \alpha\left[S_{i j}-\gamma_{i j}(S-\rho) / 2\right] \\ & - \alpha\left[S^{M}_{i j}-\gamma_{i j}(S^{M}-\rho^{M}) / 2\right], \label{subeq2}
\end{split}
\\
\partial_{\perp} \alpha =&\alpha^{2} \widetilde{\pi}-\alpha^{2} H_{T}, \label{subeq3}
\\
\partial_{t} \beta^{i} =&\beta^{j} \bar{D}_{j} \beta^{i}+\alpha^{2} \rho^{i}-\alpha D^{i} \alpha+\alpha^{2} H^{i}, \label{subeq4}
\\
\begin{split}
\partial_{\perp} \widetilde{\pi} =&-\alpha K_{i j} K^{i j}+D_{i} D^{i} \alpha+\mathcal{C}^{i} D_{i} \alpha -\kappa \alpha \mathcal{C}_{T} / 2 \\ & -4 \pi G \alpha(\rho+S) -\frac{\alpha}{2}(\rho^{M}+S^{M}),\label{subeq5}
\end{split}
\\
\begin{split}
\partial_{\perp} \rho^{i}=& \gamma^{k \ell} \bar{D}_{k} \bar{D}_{\ell} \beta^{i}+\alpha D^{i} \widetilde{\pi}-\widetilde{\pi} D^{i} \alpha-2 K^{i j} D_{j} \alpha \\ & +2 \alpha K^{j k} \Delta \Gamma_{j k}^{i}+\kappa \alpha \mathcal{C}^{i}
\\
&-16 \pi G \alpha j^{i} -2\alpha j_{M}^{i}, \label{subeq6}
\end{split}
\end{align}
 \end{subequations}
with the constraints, 
\begin{subequations} 
\label{cons_M}
\begin{align}
\mathcal{C}_{T} & \equiv \widetilde{\pi}+K, \label{const1} \\
\mathcal{C}^{i} & \equiv-\rho^{i}+\Delta \Gamma_{j k}^{i} \gamma^{j k}, \label{const2} \\ 
\mathcal{H} & \equiv K^{2}-K_{i j} K^{i j}+R-16 \pi G \rho -2\epsilon \rho^{M}, \label{const3}\\ 
\mathcal{M}_{i} & \equiv D_{j} K_{i}^{j}-D_{i} K-8 \pi G j_{i} -\epsilon j_{i}^{M}, \label{const4}
\end{align}
\end{subequations}
where $K\equiv\gamma^{ij}K_{ij}$, ${D}_{i}$ and $\bar{D}_{i}$ are 
the covariant derivatives for the three-metric $\gamma_{ij}$ and the background 3-metric  $\bar{\gamma}_{ij}$ respectively. The derivative operator $\partial_{\perp}$ is 
defined as $\partial_{\perp}=\partial_{t} - \mathcal{L}_{\beta}$, where $ \mathcal{L}_{\beta}$ is the Lie derivative along the shift vector $\beta^{i}$. Defining $\Delta 
\Gamma^{i}_{jk}:=^{(3)}\Gamma^{i}_{jk} - ^{(3)}\bar{\Gamma^{i}}_{jk}$, where these are the Christoffel symbols for the induced metric and background metric (flat in spherical coordinates) respectively. Defining also $H_{T}:= H^{a}n_{a}$, where $n_{a}$ is the normal vector to the spatial hypersurfaces defined by the spacetime foliation.
The new dynamical variables 
$\widetilde{\pi}$ and $\rho^{i}$ are introduced through equations (\ref{subeq3}-\ref{subeq4}) to make the system (ignoring the extensions to gravity) first order in time derivatives. $S_{ij}$, $S$, $\rho$ 
and $j^{i}$ are the matter variables constructed from the energy-momentum tensor $T_{a b}$ as, $S_{ij}= P^{a}_{i}P^{b}_{j} T_{a b}$, its trace $S = \gamma^{ij}S_{ij} $,  
$\rho= n_{a}n_{b}T^{a b}$, and $j^{i}=-P^{i a}n^{b}T_{a b}$. Where $P^{i a}$ is a projection tensor to the spatial hypersurface. Here the definitions for $S^{M}_{ij}$,$S^{M}$, $\rho^{M}$ and $j_{M}^{i}$  are analogous to the ones for the 
matter sources, but instead of using $T_{a b}$, we use $M_{a b}$.

 Let us now analyze the structure of the terms introduced by $M_{ab}$, which modify Einstein's equations. These terms contain up to 4th-order time and spatial derivatives of metric components. In addition, they contain nonlinear combinations of derivatives that would make the usual hyperbolicity analysis \cite{Gundlach:2004ri} inapplicable. Furthermore, the constraint equations \eqref{const3}-\eqref{const4} contain time derivatives, which are not present in the Hamiltonian and Momentum constraints in GR. These sorts of issues are not uncommon when dealing with modified gravity theories, even in Horndeski theories, which are second order in derivatives and incorporate a non-minimally coupled scalar field, suffer from pathologies that can render the problem of interest ill-posed \cite{Ripley:2019hxt, Bernard:2019fjb, Figueras:2020dzx, Thaalba:2023fmq}. In those cases, after significant theoretical efforts, appropriate new gauges were formulated \cite{Kovacs:2020pns, Kovacs:2020ywu} that ameliorate these issues to the point where nonlinear studies of compact binary mergers are possible \cite{East:2020hgw, Corman:2022xqg, AresteSalo:2022hua, East:2022rqi} for some regime of small coupling values. In the case of higher derivative extensions to GR, fully nonlinear evolution has been performed \cite{Cayuso:2023aht, Cayuso:2020lca} for an eight-dimensional operator EFT extension through controlling pathological higher frequencies via a ``fixing'' method \cite{Cayuso:2017iqc, Allwright:2018rut, Lara:2021piy, Franchini:2022ukz} leaving the long wavelength physics unaltered.

Coming back to this paper's theory of interest works like \cite{Held:2021pht, Held:2023aap} tackle these issues by re-writing the theory following the work of Noakes \cite{Noakes:1983xd}, in which the Ricci scalar and the traceless part of the Ricci tensor can be elevated to massive spin-0 and spin-2 fields and are evolved with equations derived directly from the field equations of the theory. With this prescription, they can verify numerical stability in the Ricci-flat subsector and confirm that it is indistinguishable from GR. However, an opposing view to this method can be formed from the perspective of EFT. The extra modes that this theory introduces and that this approach makes explicit have masses that are above the cut-off scale of the EFT; hence the dynamics of these modes should be irrelevant in the EFT regime\footnote{See \cite{Bueno:2023jtc} for a similar argument on the massive degrees of freedom in six-dimensional operators EFT.}. Furthermore, depending on the signs and values of $\epsilon_{1}$ and $\epsilon_{2}$, these massive degrees of freedom can become tachyonic, which would take them outside the regime of applicability of the EFT. In contrast, this work, taking this intuition from EFT, will actively remove these extra degrees of freedom by eliminating the higher order time derivatives in the field equations via an ``Order Reduction''  \cite{Solomon:2017nlh}  procedure \footnote{This ``Order Reduction'' approach is not to be confused with the ``Order reduction'' techniques used in \cite{Okounkova:2019zjf, Okounkova:2020rqw}, where order-reducing refers to replacing some problematic terms and solving them iteratively/perturbatively.}.  
Proceeding as done in \cite{Cayuso:2020lca} (see Section II-C of that work for more details), one can use the evolution and constraint equations to 0th order in $\epsilon_{1}$ and $\epsilon_{2}$ to find expressions of higher order time and spatial derivatives of the metric components in terms of lower order derivatives.

Schematically,
\begin{equation}\label{eqor}
\begin{split}
    \frac{\partial \bm{g}}{\partial t^{2}} &= \bm{E}(\bm{g},\partial_{a}\bm{g},\partial_{i}^{2}\bm{g})\\ &+\epsilon\bm{M}(\bm{g},\partial_{a}\bm{g},\partial_{a}^{2}\bm{g},\partial_{a}^{3}\bm{g},\partial_{a}^{4}\bm{g}) + \mathcal{O}(\epsilon^{2}),     
\end{split}
\end{equation}
represents the evolution system of equations \eqref{system_M} written in terms of the variables $\bm{g}= \lbrace \gamma_{ij}, \alpha, \beta \rbrace$. Here $\bm{E}$ represents the GR terms, which depend only up to first-time derivatives and second spatial derivatives of $\bm{g}$. $\bm{M}$ represents the terms from the modified theory, which depend on up to fourth-order spacetime derivatives. Truncating \eqref{eqor} to order $\mathcal{O}(\epsilon^{0})$

\begin{equation}\label{eqor0}
    \frac{\partial \bm{g}}{\partial t^{2}} = \bm{E}(\bm{g},\partial_{a}\bm{g},\partial_{i}^{2}\bm{g})+\mathcal{O}(\epsilon),     
\end{equation}
and taking derivatives of it gives expressions to higher than second-time derivatives of $\bf{g}$ in terms of lower order derivatives. This way \eqref{eqor0} and its derivatives can be used to replace $\lbrace \partial_{a}^{2}\bm{g},\partial_{a}^{3}\bm{g},\partial_{a}^{4}\bm{g}\rbrace $ in $\bm{M}$, in favor of $\widetilde{\bm{M}}$, to obtain redefinitions of \eqref{eqor} that are lower in time derivatives and valid to  $\mathcal{O}(\epsilon)$, 
\begin{equation}\label{eqorF}
\begin{split}
    \frac{\partial \bm{g}}{\partial t^{2}} &= \bm{E}(\bm{g},\partial_{a}\bm{g},\partial_{i}^{2}\bm{g})\\ &+\epsilon\widetilde{\bm{M}}(\bm{g},\partial_{a}\bm{g},\partial_{a}\partial_{i}\bm{g},\partial_{a}\partial_{i}^{2}\bm{g},\partial_{a}\partial_{i}^{3}\bm{g}) + \mathcal{O}(\epsilon^{2}),     
\end{split}
\end{equation}

This way, expressions for $S_{ij}^{M}$, $S^{M}$, $\rho^M$ and $j^{i}_{M}$, let us call them $\widetilde{S_{ij}^{M}}$, $\widetilde{S^{M}}$, $\widetilde{\rho^M}$ and $\widetilde{j^{i}_{M}}$ can be obtained, which no longer contain higher derivatives in time and that are valid to $\mathcal{O}(\epsilon_{1})$ and $\mathcal{O}(\epsilon_{2})$. Once all undesired time derivatives are eliminated, the constraint equations, which now only contain spatial derivatives, can be used to find expressions for some (not all) higher spatial derivative derivatives of the metric components in terms of lower derivatives. In spherical symmetry, even though not all higher spatial derivatives expressions are available through an order reduction of the constraints, this procedure is enough to eliminate all higher-than-second spatial derivatives of the metric components. 
During this procedure, one introduces higher-order spatial derivatives (up to third) of the scalar field $\phi$. In some way, all of the higher-order time and spatial derivatives of gravity variables have been traded for 3rd derivatives of the scalar field.
 This is seen easily by noticing that this reduction of order is equivalent to replacing $R_{a b}$ and $R$  through $T_{a b}$ in all the $\epsilon$ proportional terms in \eqref{Field_eqs_o}.
One could proceed as done in \cite{Cayuso:2023aht, Cayuso:2020lca} and control the higher frequencies via the ``fixing'' approach. One of the objectives of this work is to explore under what circumstances the system is well-behaved after performing the order reduction without attempting to control the higher frequencies.

 \section{Target problem and setup}\label{target}

The objective is to study this theory and its equations in dynamical scenarios where nonlinearities are important. We want to explore in which regime of the parameter space one can carry out numerical evolution without instabilities. If such instabilities do appear, the objective is to asses whether this happens within the regime of applicability of the EFT. To this end, we evolve spacetimes consisting of an initial in-falling scalar Gaussian profile, ultimately collapsing into a BH. This work will avoid treating critical collapse \cite{Choptuik:1992jv}, mainly because the EFT is doomed to be outside of its regime of validity during such a process.

Reducing the problem to spherical symmetry, the line element for this problem is given by,
\begin{equation} \label{ds}
\begin{split}
ds^{2}&=(-\alpha^{2}+g_{rr}\beta^{2})dt^{2} +2\beta g_{rr}drdt + g_{rr}dr^{2} \\ &
+ r^{2}g_{T}(d\theta^{2} + \sin^{2}\theta d\varphi^{2}),
\end{split}
\end{equation}
where $\alpha$ is the lapse function, $\beta$ is the radial  component of the shift vector, and $g_{rr}$ and $g_{T}$ are the radial and angular components of the spatial metric $\gamma_{ij}$.

The equations that arise from this ansatz contain factors of $r^{-p}$, which lead to divergences at the origin $r=0$. Using L'Hopital's rule, one can carefully redefine the equations at the origin to avoid these coordinate singularities. This technique is essential when dealing with the high $p$ exponents that corrections to GR introduce.

\subsection{Initial data}
 What determines whether the scalar field collapses into a BH or bounces back to infinity depends on the properties of the initial profile of the field. All of this will be encoded in the initial data prescribed. 
 In this section, we discuss how we construct initial data consistent with the constraints of the modified theory. 
 
 Starting from the conformal decomposition of the spatial metric as
  \begin{equation}
    \gamma_{ij} = \psi^{4}\tilde{\gamma}_{ij},
\end{equation}
where $\psi$ is the conformal factor and $\tilde{\gamma}_{ij}$ 
being the flat metric in spherical coordinates. With this choice, the \textit{Hamiltonian Constraint} takes the form,
\begin{equation}
 8 \nabla^{2}_{\textit{flat}}\psi + \psi^{5}(A_{ij}A^{ij}- \frac{2}{3}K^{2}) + 16\pi\psi^{5}\rho + 2\epsilon\psi^{5}\widetilde{\rho^{M}}=0, 
\end{equation}
where $A_{ij}$ is the traceless part of the extrinsic curvature tensor $K_{ij}$ and now the additional term $ 2\psi^{5}\widetilde{\rho^{M}}$ contains the modifications to GR.

The \textit{Momentum Constraint} takes the form,
\begin{equation}
 \nabla_{j}A^{ij} - \frac{2}{3}\nabla^{i}K - 8\pi j^{i} - \epsilon \widetilde{j^{i}_{M}}=0,
\end{equation}
which includes the additional current-like term $- \epsilon \widetilde{j^{i}_{M}}$. 
We take the extrinsic curvature to be traceless by setting the ansatz,
\begin{equation}
    A_{ij} = \begin{pmatrix} 	K_{rr} & 0 & 0 \\ 0 &-r^2\frac{K_{rr}}{2} & 0 \\ 0 & 0 &-r^2\frac{K_{rr} \sin^{2}\theta}{2}  \end{pmatrix}.
\end{equation}
The expressions of the Hamiltonian and Momentum constraint under such ansatz read,
\begin{equation}\label{ini_eq_psi}
\begin{split}
 \frac{\partial^{2} \psi}{\partial r^{2}} & = -\frac{2}{r}\frac{\partial\psi}{\partial r}- \frac{3}{16}\frac{K_{rr}^{2}}{\psi^{3}}  - \pi \psi \left(\frac{\partial \phi}{\partial r}\right)^{2} 
 -\pi\psi^{5}\Sigma^{2} \\ 
 & -\frac{1}{4}\epsilon\psi^{5}\widetilde{\rho^{M}}, 
\end{split} 
\end{equation}
\begin{equation}\label{ini_eq_Krr}
\begin{split}
 \frac{\partial K_{rr}}{\partial r} &= -2\psi^{-1}K_{rr}\frac{\partial \psi}{\partial r} - \frac{3}{r}K_{rr} 
 + 8\pi \psi^{4}\Sigma\frac{\partial \phi}{\partial r } \\
 & + \epsilon \psi^{8}\widetilde{j^{r}_{M}}.
\end{split} 
\end{equation}

 Notice that $\widetilde{\rho^{M}}$ and $\widetilde{j^{i}_{M}}$ are the order reduced expressions that we obtained after the order reduction procedure, and when evaluated under this ansatz possess only up to first order derivatives of $\psi$ and no derivatives of $K_{rr}$. In this form, these equations can be integrated directly to find solutions once the scalar field initial data is specified and appropriate boundary conditions set. This technique was used in \cite{Cayuso:2020lca}, as ``order-reduced direct integration'', to successfully construct BH initial data in spacetimes in the presence of a scalar field for an eight-dimensional operator EFT of GR.

\subsubsection{Scalar field}
The initial scalar field is prescribed such that it is initially mostly in-falling towards the origin; this can be achieved by having a field of the form,
\begin{equation}
 \phi(t,r) = \frac{\Phi( u )}{r},
\end{equation}
where $u \equiv r +t$ and,  
\begin{equation}
\Phi(u) = Au^{2}\exp\left(-\frac{(u-r_{c})^{2}}{\sigma^{2}}\right),
\end{equation}
where $A, r_c$ and $\sigma$ are the amplitude, center, and width of the pulse respectively.
Under this choice, the initial values of  scalar field variables are given by,
\begin{equation}
 \phi_{0}\equiv \phi(t=0,r) =Ar\exp\left(-\frac{(r-r_{c})^{2}}{\sigma^{2}}\right),
\end{equation}
\begin{equation}
 \Sigma(t=0,r)=\frac{\phi_{0}}{\alpha}\left( \beta\left(\frac{1}{r} - \frac{2(r-r_{c})}{\sigma^{2}} \right) 
 - \left( \frac{2}{r} - \frac{2(r-r_{c})}{\sigma^{2}}\right)\right),
\end{equation}

where $\Sigma$ is defined as,
\begin{equation}
    \Sigma(t,r) = \frac{1}{\alpha}\left(\beta\frac{\partial \phi}{\partial_{r}} -\frac{\partial \phi}{\partial_{t}}\right).
\end{equation}

\subsubsection{Boundary conditions}
To construct the initial data, boundary conditions for the fields must be prescribed.  Regularity at the origin imposes $\Omega(r=0)\equiv \partial_{r}\psi(r=0) = 0$. For convenience, we can set $K_{rr} = 0$ at the origin. To determine the remaining condition on the $\psi$ field we impose that the exterior boundary conditions should have the following form,

\begin{align}
 &\left. \psi\right|_{r_{out}}= 1 + \frac{M}{2 r_{out}}, \label{psi_rin} \\
 & \left.\frac{\partial \psi}{\partial r}\right|_{r_{out}} = -\frac{M}{2 r_{out}^{2}}, \label{dpsi_rin}
\end{align}
where $r_{out}$ is the exterior grid boundary and $M$ is the ADM mass (which will depend on the scalar field initial configuration). A way to achieve this is to perform a shooting procedure on the value of $\psi(r=0)$ such that the integrated solution on the outer boundary satisfies,

\begin{equation}
    \left. \psi\right|_{r_{out}}= 1 - r_{out}\left.\frac{\partial \psi}{\partial r}\right|_{r_{out}},
\end{equation}
we achieve this by implementing a Newton-Raphson method.

We impose that the initial values of gauge variables satisfy, 
\begin{align}
 &\alpha(t=0) = 1, \\
 &\beta(t=0) = 0, \\
 &\widetilde{\pi}(t=0) = 0,\\
 &\rho^{i}(t=0) = -2\psi^{-5}\Omega,
\end{align}
where the last two are required to initially satisfy the constraints \eqref{const1}-\eqref{const2}.

\subsection{Numerical implementation}
The following numerical scheme is implemented to evolve the system presented in section \ref{evolution_equations}. Time is integrated through a 4th-order Runge-Kutta with a CFL coefficient such that $dt=0.25 dx$, where $dt$ is the time-step and $dx$ denotes the uniform spatial grid spacing. Spatial derivatives are discretized via Finite Differences operators, which are 6th-order accurate in the interior and 3rd-order in the boundaries. Kreiss-Oliger dissipation is implemented with operators that are 8th-order accurate in the interior and 4th-order in the boundary. 
When no BH is present in the simulation, the grid extends from $r_{i}=0$ to $r_{out}= 200$. During the evolution, the appearance of an apparent horizon is monitored; if one appears, then the code will excise a portion (including $r=0$) of the domain contained inside this apparent horizon. 
A damped harmonic gauge \cite{Choptuik:2009ww, Lindblom:2009tu, Varma:2018evz} is adopted, which sets the gauge source vector to satisfy:  $H_{a} = z(\log{(\sqrt{g_{rr}}g_{T}\alpha^{-1})}n_{a} - g_{a b}\beta^{b}\alpha^{-1})$. We take a fixed value of $z=0.5$. 

\subsection{Monitoring quantities}

As previously mentioned, an EFT description of a system involves a truncated expansion of a tower of curvature operators, and control over this expansion is lost if the curvature becomes too large. Determining whether the system remains within the regime of applicability of the EFT throughout evolution is a necessary condition \footnote{Even if the theory is at all times within the EFT's regime of validity, undesired issues such as secular effects \cite{Okounkova:2019zjf, Okounkova:2020rqw} could emerge and spoil the physics.} to guarantee that the observed behavior is representative of the true physics of the underlying theory in the low energy regime.

A reasonable indicator of whether the system is within the regime of applicability of the EFT is to compare if terms that are higher order in the perturbation scheme remain subdominant to lower order ones \cite{deRham:2020zyh, Chen:2021bvg}. For example one expects that $\mid R \mid > \mid  \epsilon_{1}R_{ab}R^{ab} \mid + \mid  \epsilon_{2}R^2 \mid$. Using the fact that  $R_{ab} = 8\pi( T_{a b} -1/2 T g_{ab}) + \mathcal{O} (\epsilon_1,\epsilon_{2})$, (ignoring higher order terms in $\epsilon_{1}$ and $\epsilon_{2}$) the inequality can be expressed as:
\begin{equation}\label{EFTcondition_Ricci}
 \mathcal{E}_{R} \equiv  8\pi (\mid\epsilon_{1}\mid +\mid\epsilon_{2}\mid) \mid\left(-\Sigma^{2}g_{rr} + (\partial_{r}\phi)^2 \right) \mid  g_{rr}^{-1} <1.
\end{equation}

Another indicator that can be used to discern whether the theory remains in the EFT regime of applicability is through some curvature invariant that is non-vanishing for vacuum spacetimes, for instance, the Kretschmann scalar $\mathcal{C}\equiv R_{a b c d}R^{a b c d}$. Using this invariant, a natural threshold for the regime of applicability of the EFT is given by $\Lambda^{-2}\mathcal{C} > \Lambda^{-6}\mathcal{C}^{2} $, which can be easily rewritten as,

\begin{equation}\label{EFTcondition_Kret}
    \mathcal{E}_{\mathcal{C}} \equiv \mathcal{C}\Lambda^{-4} \approx \mathcal{C}\max(\epsilon_{1}^{2},\epsilon_{2}^{2}) < 1.
\end{equation}

During evolution, these two quantities will be monitored to get an idea whether the system is in the validity regime of the EFT, close to leaving it or outside of it\footnote{There are, of course, many other quantities one could check, for example, checking that the six-dimensional operators should be subdominant to the four-dimensional ones, for example, $ R_{ab}R^{ab}\Lambda^{-2} > R_{a b }^{\,\,\,\,\,\,ef}R^{a b c d}R_{cdef}\Lambda^{-4}$.}. 

The way the equations have been rewritten after the order reduction is now somewhat more familiar to the equations we might encounter in GR, where metric components appear at most as second derivatives, and these second derivatives appear linearly in the equations.
The system's characteristic speeds are usually evaluated to study hyperbolicity and, consequently, the well-posedness of an initial value problem. However, the presence of the third-order derivative of the scalar field in the gravitational equations prevents us from carrying out this analysis. One can, however, attempt to get some insight out of that procedure by computing the characteristic speeds by considering solely the gravitational sector (\ref{subeq3}-\ref{subeq6}) and considering the $\phi$ field as a source\footnote{Note that $\phi$ evolves with $\Box\phi=0$, so in a very local sense, its evolution should be well-posed}. The characteristic matrix of that system is diagonalizable and possesses the following eigenvalues (characteristic speeds):

\begin{subequations}\label{eigenvalues}
\begin{align}
\lambda_{1\pm} &= \beta \pm \frac{\alpha}{\sqrt{g_{rr}}}, \\
\lambda_{2\pm} &= \beta \pm \frac{\alpha\sqrt{8\pi(2\epsilon_{1} -4\epsilon_{2})(\partial_{r}\phi)^{2}  +g_{rr}}}{g_{rr}}, \\
\lambda_{3\pm} &= \beta \pm \frac{\alpha\sqrt{8\pi(4\epsilon_{1} -8\epsilon_{2})(\Sigma^{2}g_{rr}-(\partial_{r}\phi)^{2})  +g_{rr}}}{g_{rr}}.
\end{align}    
\end{subequations}

Notice how all velocities in \eqref{eigenvalues} reduce to what one obtains in GR when $\phi=0$, when $\epsilon_{1}=\epsilon_{2}=0$ or when $\epsilon_{1}=2\epsilon_{2}$. While the two first conditions imply that the theory reduces to GR, for the last one $\epsilon_{1}=2\epsilon_{2}$, the equations are still different from GR. When neither of those conditions is met, these speeds  are modified from the GR ones and are real only under certain conditions. The radicand on $\lambda_{2\pm}$:
\begin{equation}
\chi_{2} \equiv 8\pi(2\epsilon_{1} -4\epsilon_{2})(\partial_{r}\phi)^{2}  +g_{rr},
\end{equation}
can become negative if $\epsilon_{1} -2\epsilon_{2}<0$ and $g_{rr}<-8\pi(2\epsilon_{1} -4\epsilon_{2})(\partial_{r}\phi)^{2}$ which is possible if the scalar field gradients are large enough. Furthermore, regardless of the sign of $(\epsilon_{1} -2\epsilon_{2})$ the radicand of $\lambda_{3\pm}$:

\begin{equation}\label{eq:chi3}
\chi_{3}\equiv 8\pi(4\epsilon_{1} -8\epsilon_{2})(\Sigma^{2}g_{rr}-(\partial_{r}\phi)^{2})  +g_{rr},
\end{equation}

can become negative, the factor $(\Sigma^{2}g_{rr}-(\partial_{r}\phi)^{2})$ does not have definite sign and for large enough $\Sigma$ or $\partial_{r}\phi$ then  $\chi_{3}<0$ is a possibility.

This (simplified) analysis tells us that the system could undergo a character transition \cite{Bernard:2019fjb} during evolution rendering the problem ill-posed. The appearance of this transition could depend on the initial data prescribed; for example, a collapsing field would evolve to have very large gradients and trigger this transition, while a different configuration could avoid it. In this work, we will explore the evolution of a collapsing scalar field for different values of the coupling parameters and try to identify if such a transition happens, whether it triggers instabilities, and whether it occurs inside of the regime of applicability of the EFT. 

\section{Results}\label{results}

We turn our attention now to the evolution of the in-falling self-gravitating scalar field with different choices of the coupling parameters $\lbrace \epsilon_{1}, \epsilon_{2}\rbrace$. Whether the incoming pulse collapses into a BH or bounces back to infinity will depend mostly on the choice of its initial parameters, amplitude $A$, width, $\sigma$, and position $r_{c}$. To study the collapse case, these three parameters will be fixed to $A=0.0023$, $\sigma= 1$, and $r_c = 10$. For these values in the initial scalar profile, the ADM mass of the system is $M_{ADM} = 1.024$ when $\epsilon_{1}=\epsilon_{2}=0$. The relevant length scales in the modified theory ($|\epsilon_{1}|^{1/2} \approx |\epsilon_{2}|^{1/2} \approx \Lambda^{-1} $) should be then compared to the mass of the system. For reference, when these couplings are large $|\epsilon_{1}|=|\epsilon_{2}|=0.1$ the difference in $M_{ADM}$ is at the sub-percent level. Even though this work focuses on the collapsing scenario, the non-collapsing scenario was also studied. The evolution of that scenario in the regime of couplings explored is well-behaved up to $|\epsilon| \approx 10^{-1}$. Above such couplings,  the system leaves the regime of applicability of the EFT. The evolution of the collapse scenario is more interesting, as we shall see in this section.

The main objective of these simulations is to explore how the evolution is altered as we modify the coupling parameters $\lbrace \epsilon_{1}, \epsilon_{2}\rbrace$, such as the behavior of the apparent horizon and curvature invariants. When couplings are turned off, and GR is evolved, the initial pulse propagates toward the origin until a BH forms. It quickly accretes the scalar field and settles to its final configuration. The final mass of the formed BH is of $M_{BH}\approx1.022$, indicating that only a very small portion of the scalar field is not accreted by the BH. To study how this same scenario would evolve when couplings are non-vanishing, an array of simulations is run with pairs of $\epsilon_{1}$ and $\epsilon_{2}$ taking values from  $\lbrace0, \tilde{\epsilon}_{n\pm}\rbrace$, with  $\tilde{\epsilon}_{n\pm} = \pm\epsilon_{0}2^{n}$ for $n=0..11$, with $\epsilon_{0}=10^{-4}$.

\begin{figure}[t!]
\includegraphics[width=0.45\textwidth]{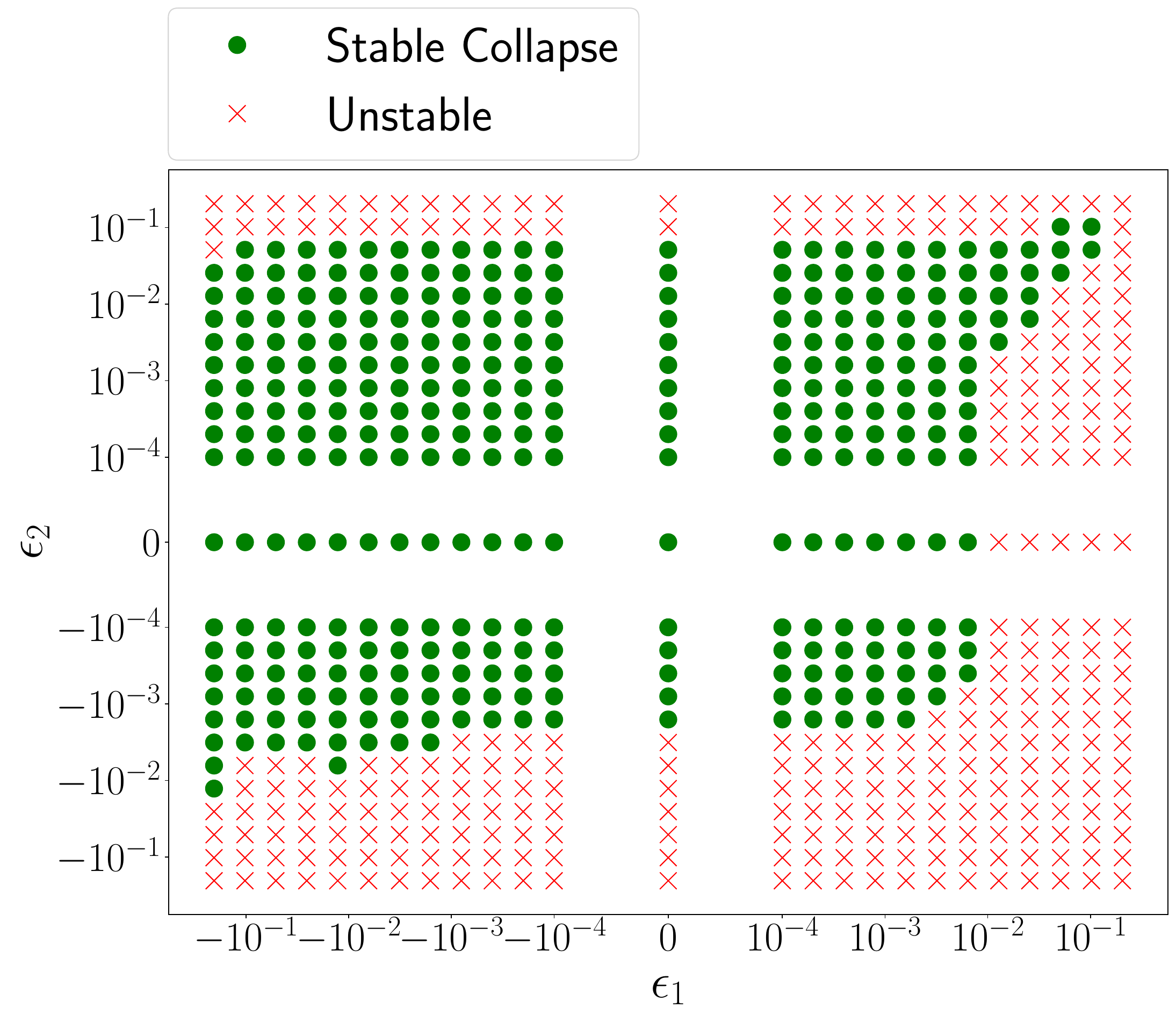}
\caption{Parameter space of simulations for the collapse scenario. In green dots are simulations that are stable  and collapse into a BH, and in red crosses are simulations that develop instabilities and crash.} 
\label{fig:Collapse_vs_chrash}
\end{figure}

Figure \ref{fig:Collapse_vs_chrash} displays whether the evolution for a pair of values $\lbrace\epsilon_{1},\epsilon_{2}\rbrace$ is stable and collapses into a BH (green dots) or if it develops instabilities and crashes (red crosses). This figure shows that there are a lot of points in the parameter space which develop instabilities, mostly when at least one of the couplings is large, especially for large and positive $\epsilon_{1}$ and large and negative $\epsilon_{2}$. 

\begin{figure}[t!]
\includegraphics[width=0.45\textwidth]{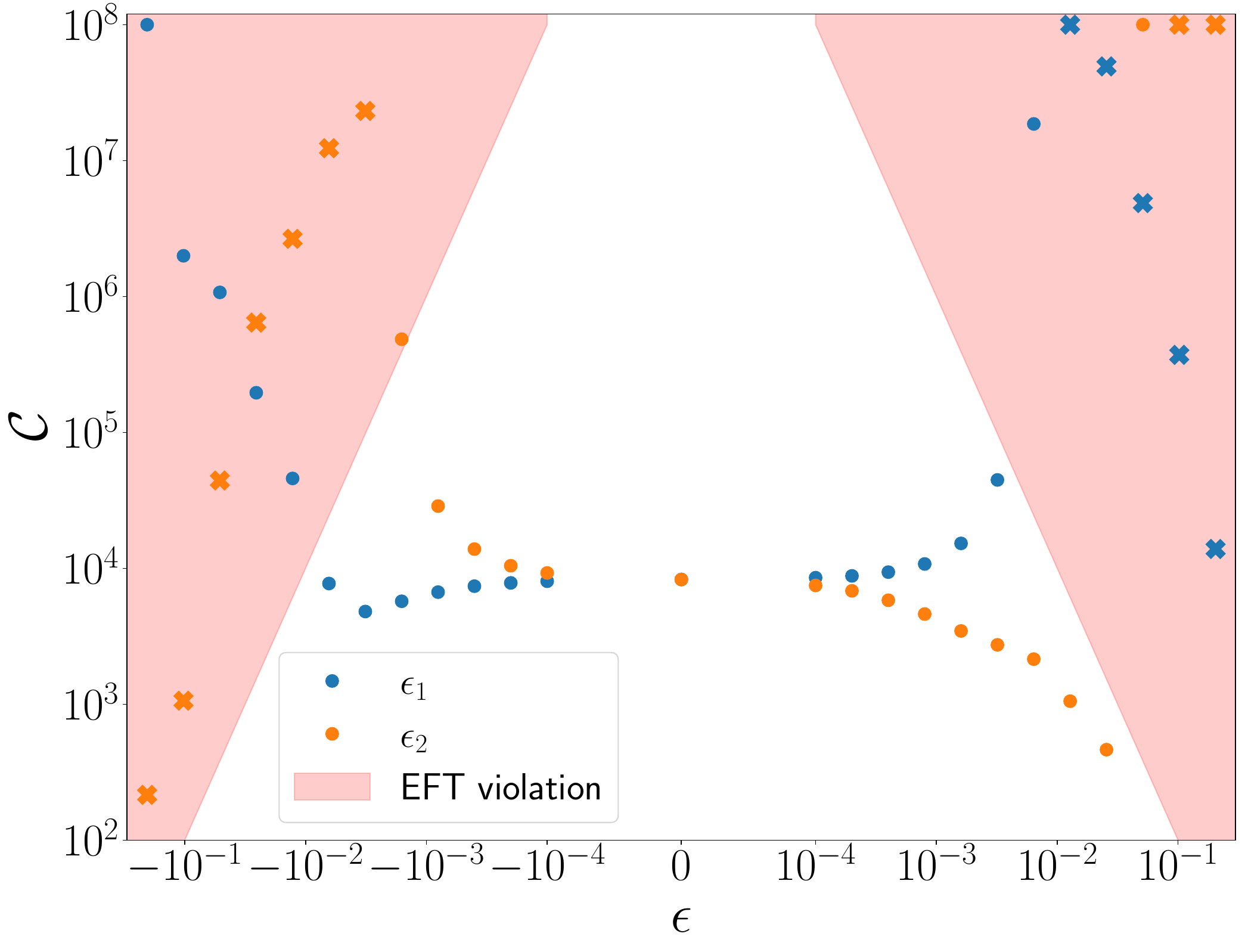}
\caption{Maximum value of $\mathcal{C}$ across space and time for simulations in the collapse scenario for either $\epsilon_{1}\neq 0$ or $\epsilon_{2}\neq 0$. Dots indicate simulations that collapsed into BHs and remained stable; crosses indicate simulations that crashed. The red shaded region indicates values of $\mathcal{C}$ that lie outside of the regime of applicability of the EFT in accordance with \eqref{EFTcondition_Kret}. Values of $\mathcal{C} > 10^8$ have been labeled as $10^8$ for convenience. } 
\label{fig:kretschmann_limits}
\end{figure}

\begin{figure*}
  \includegraphics[width=\textwidth,height=5.5cm]{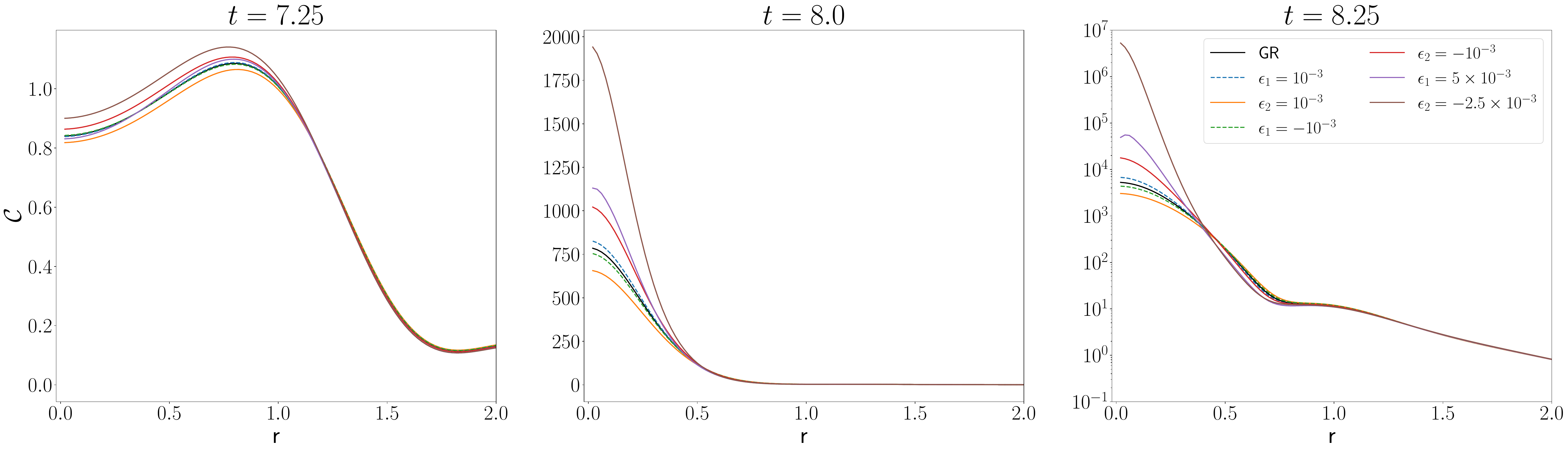}
  \caption{Snapshots of radial profiles of $\mathcal{C}$ at different times close to the collapse into a BH for different values of $\epsilon_{1}$ and $\epsilon_{2}$.}
\label{fig:c_profiles}  
\end{figure*}

To better understand what is happening, we will first focus on simulations with either $\epsilon_{1} = 0$ or $\epsilon_{2} = 0$ to study those terms individually. In Figure  \ref{fig:kretschmann_limits}, we plot the maximum value of the Kretschmann scalar $\mathcal{C}$ in space and time for this subset of the parameter space. Here dots represent simulations that were stable during the evolution and collapsed into BHs, while the crosses represent simulations that crashed. This figure shows how, relative to GR, a positive(negative) value of $\epsilon_{1}$($\epsilon_{2}$) tends to amplify the maximum value of $\mathcal{C}$ achieved during the evolution. Similarly (for small enough) negative(positive) values of $\epsilon_{1}$($\epsilon_{2}$) induce a suppression on the maximum value of $\mathcal{C}$. The magnitude of these amplifications or suppression grows as the scalar pulse approaches the origin, and corrections to GR become stronger. In Figure \ref{fig:c_profiles}, we plot several snapshots of the $\mathcal{C}$ radial profile close to the collapse to a BH. Notice however in Figure \ref{fig:kretschmann_limits} how for $\epsilon_{1} \lessapprox -10^{-2}$ the behavior of $\mathcal{C}$ drastically changes to amplification as opposed to suppression.

An indicator that the evolution for $\epsilon_{1} \lessapprox -10^{-2}$ is pathological and not physical is its convergence, which we display in Figure \ref{fig:convergence}.  This figure shows how convergence falls rapidly as the scalar field approaches the origin in cases with $\epsilon_{1}<0$ especially losing all convergence for cases with $\epsilon_{1} \lessapprox -10^{-2}$. Furthermore, one can see that the constraints in this regime of the couplings, as shown in Figure \ref{fig:constraints}, show violations above the one percent level, which indicates one should question the validity of the results.

Figure \ref{fig:kretschmann_limits} also shows in the red shaded region the values of the Kretschmann scalar $\mathcal{C}$ that would violate the EFT limit for each value of $\epsilon$ in accordance with \eqref{EFTcondition_Kret}. Interestingly a small negative value of $\epsilon_{1}$ shows a suppression of $\mathcal{C}$, which in principle, helps to avoid the restricted region. However, as $\epsilon_{1}$ becomes more negative at some point, an instability is triggered, generating an amplification of $\mathcal{C}$, clearly driving the system outside of the EFT regime of applicability. Here it is important to stress the order of these events. If an instability was generated once the system was already outside the EFT regime, this means that physics drove the system there and not pathologies. Suppose the system naturally explores higher curvatures and numerical instabilities appear after leaving the regime in which the EFT approach is valid. In that case, we need not worry about these simulations crashing and acknowledge the inadequacy of the EFT prescription to describe these scenarios. This seems to be the case for positive(negative) values of $\epsilon_{1}$($\epsilon_{2}$), which induce an amplification on $\mathcal{C}$ which drives the system outside of the valid EFT regime for $\mid\epsilon\mid  \gtrapprox 10^{-3} $ and crash. In contrast, positive values of $\epsilon_{2}$ which induce suppression of $\mathcal{C}$ manage to stay within the regime of applicability of the EFT and stable up to values of $\epsilon_{2} \lessapprox 5\times 10^{-2}$, beyond this values some instabilities are triggered, the system leaves the regime of applicability of the EFT and crashes. Both large negative values of $\epsilon_{1}$ and large positive values of $\epsilon_{2}$ seem to be developing instabilities when they are within the regime of applicability of the EFT. Perhaps for these regimes, controlling the higher frequencies via a ``fixing'' approach as in \cite{Cayuso:2023aht, Cayuso:2020lca} could result in the resolution of the instabilities, but this is outside the scope of this work.

\begin{figure}[t!]
\includegraphics[width=0.45\textwidth]{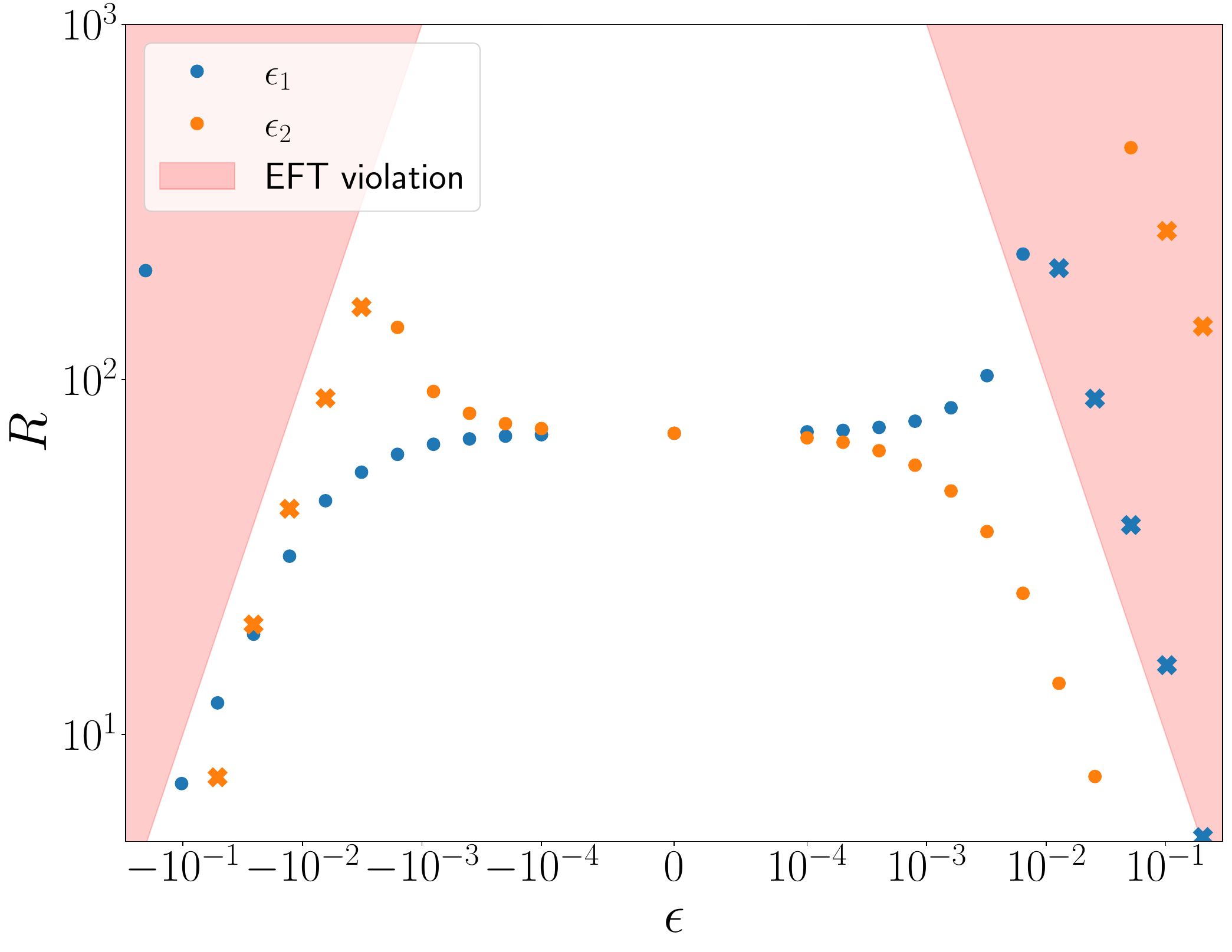}
\caption{Maximum value of the Ricci scalar $R$ across space and time for simulations in the collapse scenario for either $\epsilon_{1}\neq 0$ or $\epsilon_{2}\neq 0$. Dots indicate simulations that collapsed into BHs and remained stable; crosses indicate simulations that crashed. The red shaded region indicates values of $R$ that lie outside of the regime of applicability of the EFT in accordance with \eqref{EFTcondition_Ricci}.} 
\label{fig:Ricci_limits}
\end{figure}

Similar behavior is observed on the maximum value of the Ricci scalar $R$, which we show in Figure \ref{fig:Ricci_limits}, where we also include the EFT of applicability exclusion region in shaded red as indicated by the relation $\mathcal{E}_{\mathcal{R}}<1$, see eq.\eqref{EFTcondition_Ricci}. Interestingly, all of the simulations that crashed for $\epsilon_{2}<0$ do so within the allowed EFT regime dictated by \eqref{EFTcondition_Ricci}; however, they are outside of the valid regime according to \eqref{EFTcondition_Kret}.

\begin{figure}[t!]
\includegraphics[width=0.45\textwidth]{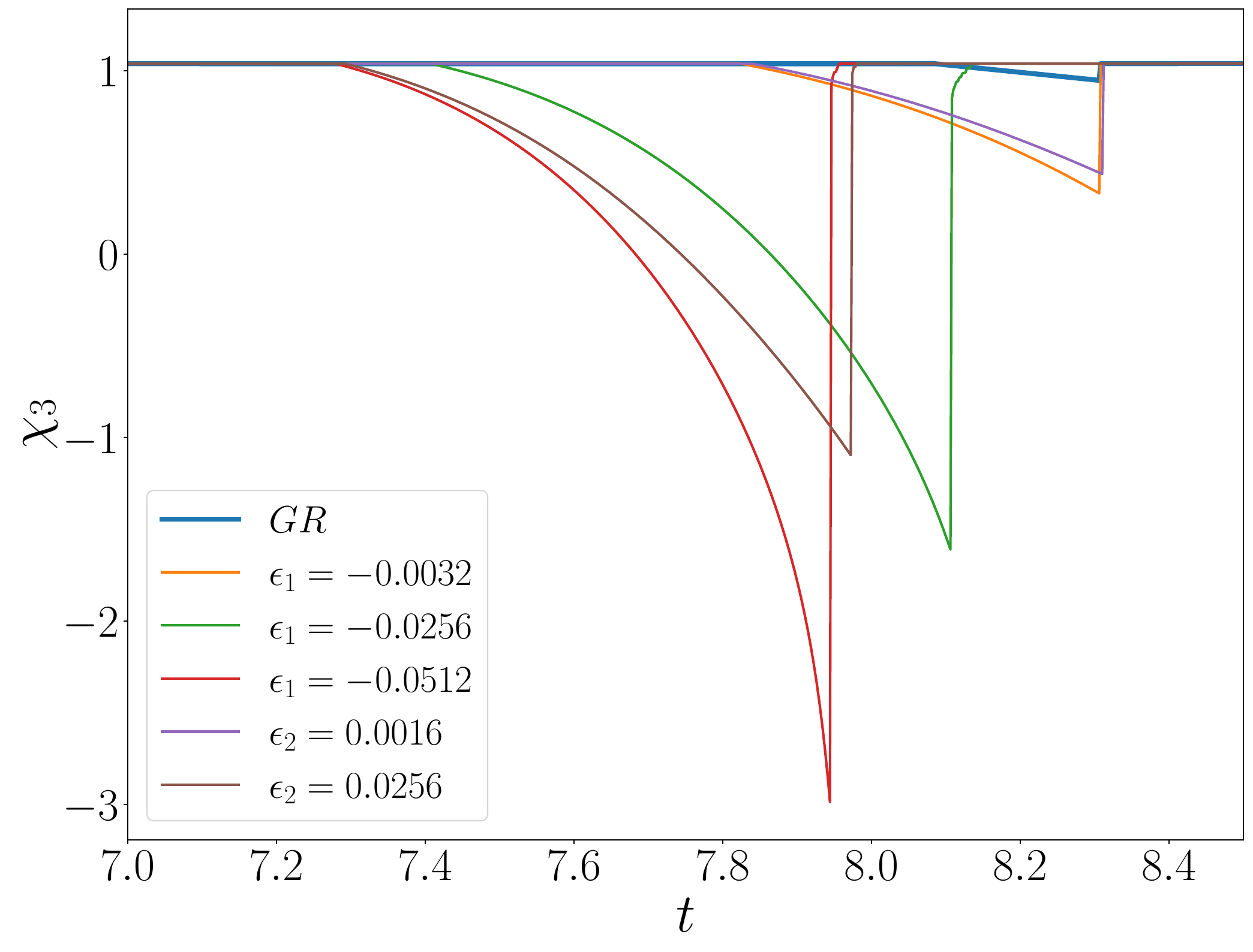}
\caption{Minimum value of $\chi_{3}$ (the radicand of the eigenvalue $\lambda_{3\pm}$)  as a function of time for simulations in the collapse case for different values of $\epsilon_{1}$ and $\epsilon_{2}$. Once an apparent horizon is found, the minimum is computed outside the horizon, hiding negative values inside; this explains the sharp transitions.} 
\label{fig:X3}
\end{figure}

Another quantity that we can inspect is the radicand $\chi_{3}$, see eq.\eqref{eq:chi3}, of the eigenvalue $\lambda_{3\pm}$, which, as we stated before, if it becomes negative could be related to a character transition and the breakdown of the initial value problem. Figure \ref{fig:X3} shows the spatial minimum  value of $\chi_{3}$ as a function of time for simulations $\epsilon_{1}<0$ or $\epsilon_{2}>0$, which are the cases in which $\chi_{3}$ decreases towards 0 and negative values. As Figure \ref{fig:X3} shows for small(large) enough values of $\epsilon_{1}$($\epsilon_{2}$) $\chi_{3}$ can become negative. As mentioned, very negative values of $\epsilon_{1}$ trigger instabilities, losing convergence and leaving the EFT's applicability regime. Similar issues are present for large positive values of $\epsilon_{2}$ where also $\chi_{3}<0$. However, such issues manifest before the $\chi_{3}<0$ threshold is violated. This suggests that this violation might not be the root cause of the instabilities but rather serve as a reliable indicator of their presence. This is not unexpected since this condition was built from an incomplete characteristic analysis in which the scalar field was considered a source, ignoring the presence of the higher derivatives of the field in the gravitational equations.

\begin{figure}[t!]
\includegraphics[width=0.45\textwidth]{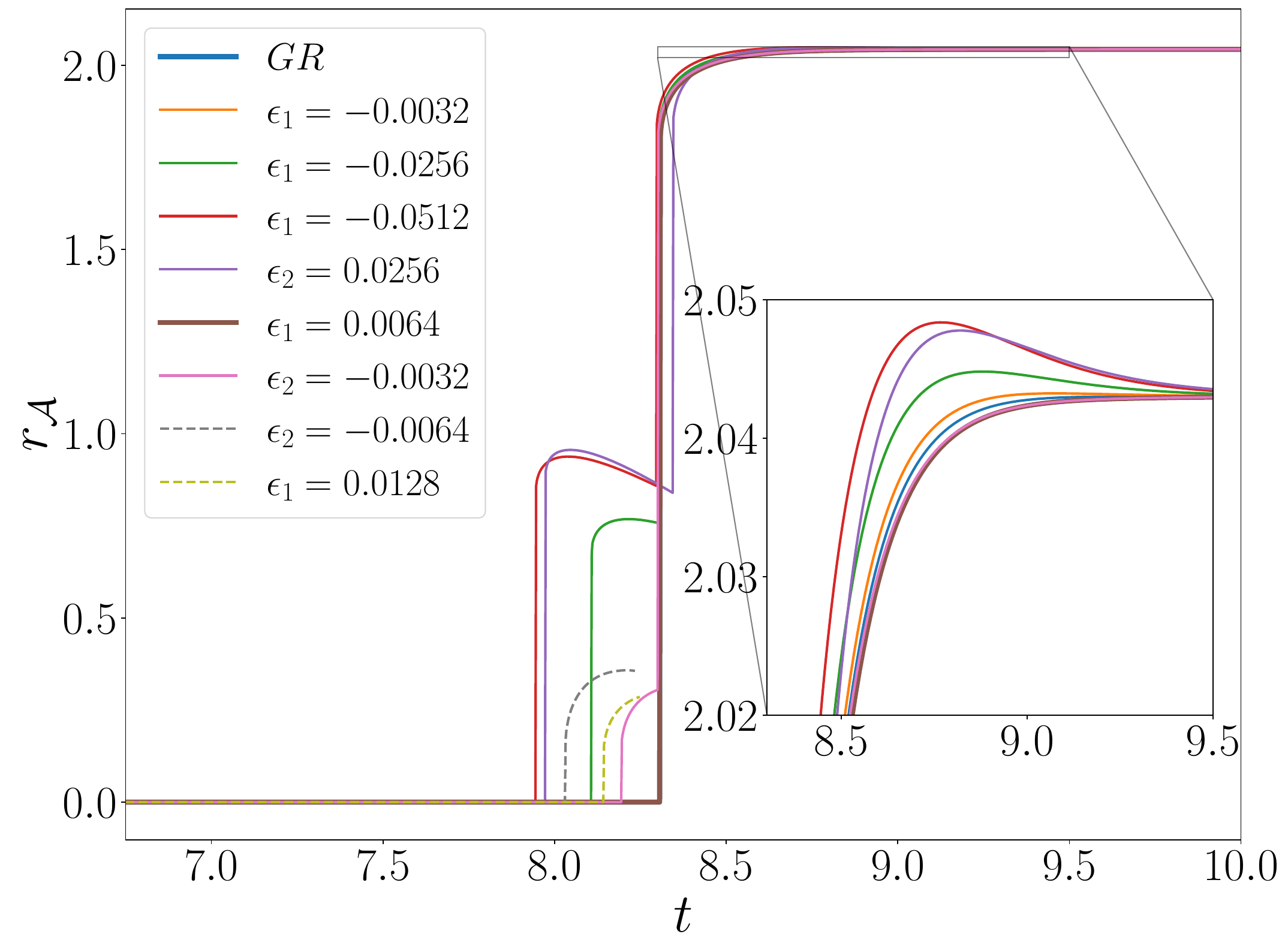}
\caption{Areal radius $r_\mathcal{A}$ of the apparent horizon as a function of time for different values of $\epsilon_{1}$ and $\epsilon_{2}$. The dashed curves correspond to simulations that crashed after the appearance of the apparent horizon. } 
\label{fig:horizons_eps1}
\end{figure}

A noticeable effect that can be appreciated in Figure \ref{fig:X3} is that simulations that develop negative values of $\chi_{3}$ also form an apparent horizon sooner than the $\chi_{3}>0$ or GR cases. Figure \ref{fig:horizons_eps1} shows the areal radius $r_{\mathcal{A}}$ of the formed horizons as a function of time for different values of the couplings. The behavior for the GR case is as expected; around $t\approx 8.3$, an apparent horizon is found, and the areal radius quickly grows until all the scalar profile has been accreted and then relaxes to its final state. This is the same behavior that some of the curves in the plot, for example, for $\epsilon_{1}=-0.0032$, $\epsilon_{1}=0.0064$, with the only difference that these curves follow slightly above and below the GR curve respectively. In contrast, for the  $\epsilon_{1}=-0.0256$, $\epsilon_{1}=-0.0512$, $\epsilon_{2}=0.0256$ cases, also shown in Figure   \ref{fig:horizons_eps1}, the systems experience  premature collapses to smaller BHs, after that $r_{\mathcal{A}}$ undergoes a brief growth, and then a substantial decrease before a new larger horizon (roughly the same size of the GR horizon) is formed. At this stage, we can see how the $r_{\mathcal{A}}$ grows above the GR curve before decreasing\footnote{ The decrease of the BH's areal radius, and hence, decrease of its area is related to violations of the Null Convergence Condition \cite{Hawking:1973uf, Hayward:1993mw}, similar behavior was observed in \cite{Cayuso:2020lca}} to join it as the final BH relaxes. Figure \ref{fig:horizons_eps1} also shows in dotted lines ($\epsilon_{1}=0.0128$ and $\epsilon_{2}=-0.0064$ ) a couple of simulations that crashed, these also display the premature appearance of a small horizon before crashing. It is important to note that all of the simulations that show this type of exotic horizon behavior evolve away from the regime of applicability of the EFT defined by \eqref{EFTcondition_Kret}. The late-time behavior of all simulations, as shown in the plot, is similar; the final BH in all cases is essentially the same. This is not unexpected; once the scalar field has been accreted by the BH and the spacetime is essentially vacuum, the equations \eqref{Field_eqs_o} reduce to Einstein's equation and can be evolved for very long times.
 
Having studied the $\epsilon_{1}$ and $\epsilon_{2}$ cases individually, we can outline a few observations.
\begin{enumerate}
    \item Positive(negative) values of $\epsilon_{1}$($\epsilon_{2}$) strongly amplify the maximum value of curvature invariants such as $\mathcal{C}$ and $R$ in contrast to GR. Their simulations are well behaved as long as the system stays within the regime of applicability of the EFT stipulated by \eqref{EFTcondition_Ricci}-\eqref{EFTcondition_Kret}, beyond that regime simulations tend to crash.
    \item Negative(positive) values of $\epsilon_{1}$($\epsilon_{2}$) strongly suppress the maximum value of curvature invariants such as $\mathcal{C}$ and $R$ in contrast to GR. Even though the suppression of these curvature invariants would help keep the system within the regime of applicability of the EFT, for large enough values of the coupling (especially for $\epsilon_{1}$), the solutions lose convergence, and the suppression becomes an amplification, driving the system outside of the EFT regime.
    \item When the couplings are sufficiently small and within the regime of the EFT, the behavior of the BH formed is very similar to that of the BH formed in the GR case. Once the horizon is formed, the high curvature regions are hidden past the horizon, making modifications extremely small.
    \item When the couplings are large enough, the BH formation becomes more exotic. Premature smaller BHs can form before a horizon similar to the one formed in the GR case appears. In addition, these smaller BHs can shrink in size during their short existence. Note, however, that the simulations in these regimes are always outside of the regime of applicability of the EFT, and hence the relevance of these results should be questioned. 
\end{enumerate}

\begin{figure}[t!]
\includegraphics[width=0.45\textwidth]{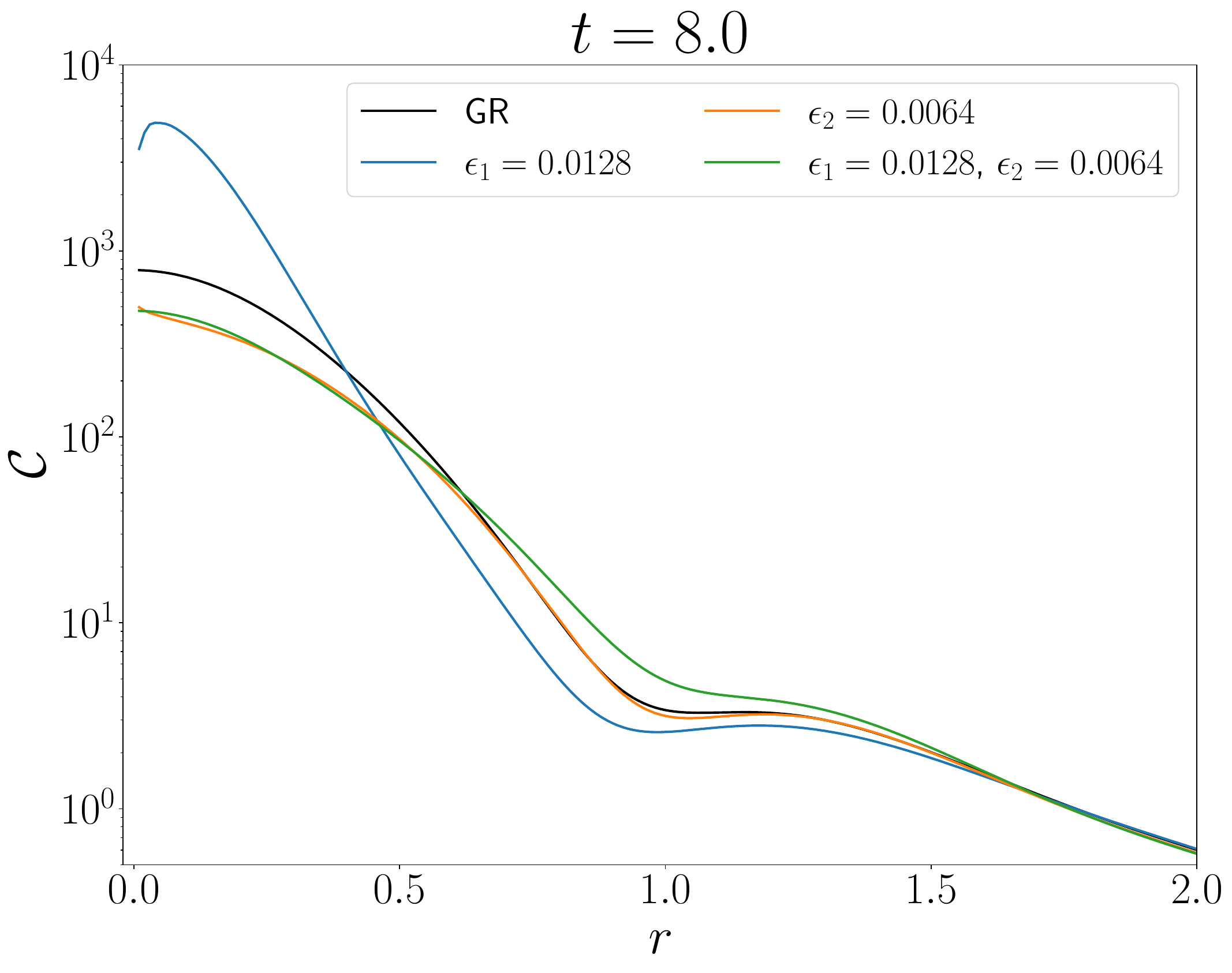}
\caption{Snapshot of radial profile of $\mathcal{C}$ at t=8.11 for simulations  with pairs of values of $\epsilon_{1}$ and $\epsilon_{2}$. Notice how the simulation with $\epsilon_{1}=0.0128$ and $\epsilon_{2}=0.064$ does not achieve the large values of $\mathcal{C}$ that the simulation with only $\epsilon_{1}=0.0128$ does.} 
\label{fig:C_both}
\end{figure}

With these observations, the interpretation of results where both $\epsilon_{1}$ and $\epsilon_{2}$ are non-zero is more direct. With our definitions of $\epsilon_{1}=a_{1}\Lambda^{-2}$ and $\epsilon_{2}=a_{2}\Lambda^{-2}$, $\Lambda$ has dimension of inverse length and both $a_{1}$ and $a_{2}$ are dimensionless. For the most part, when one of the couplings is large, and the other small, the behavior of the system is closer to the behavior of the large coupling, as we observed in the $\epsilon_{1}\neq0$ or $\epsilon_{2}\neq0$. More interesting behavior is observed when $\epsilon_{1}$ and $\epsilon_{2}$ are of the same order. For example, in the case where both $\epsilon_{1}$ and $\epsilon_{2}$ are positive, there is a competition between suppression and amplification induced in the curvature invariants, sometimes allowing the system to evolve with larger values of these couplings (in comparison to the individual cases) and stay within the regime of applicability of the EFT. This is the case for simulations with $\epsilon_{1} \approx 2\epsilon_{2}$ as it can be seen in Figure \ref{fig:C_both} were a snapshot of the radial profile for $\mathcal{C}$ is plotted in such configurations. In the case where the signs of the couplings are opposite, the effects of their terms tend to push in the same direction and consequently sometimes take the system outside of the valid regime or trigger instabilities at smaller values of the coupling in comparison to the individual $\epsilon_{1}$ or $\epsilon_{2}$ cases. 

\begin{figure}[t!]
\includegraphics[width=0.45\textwidth]{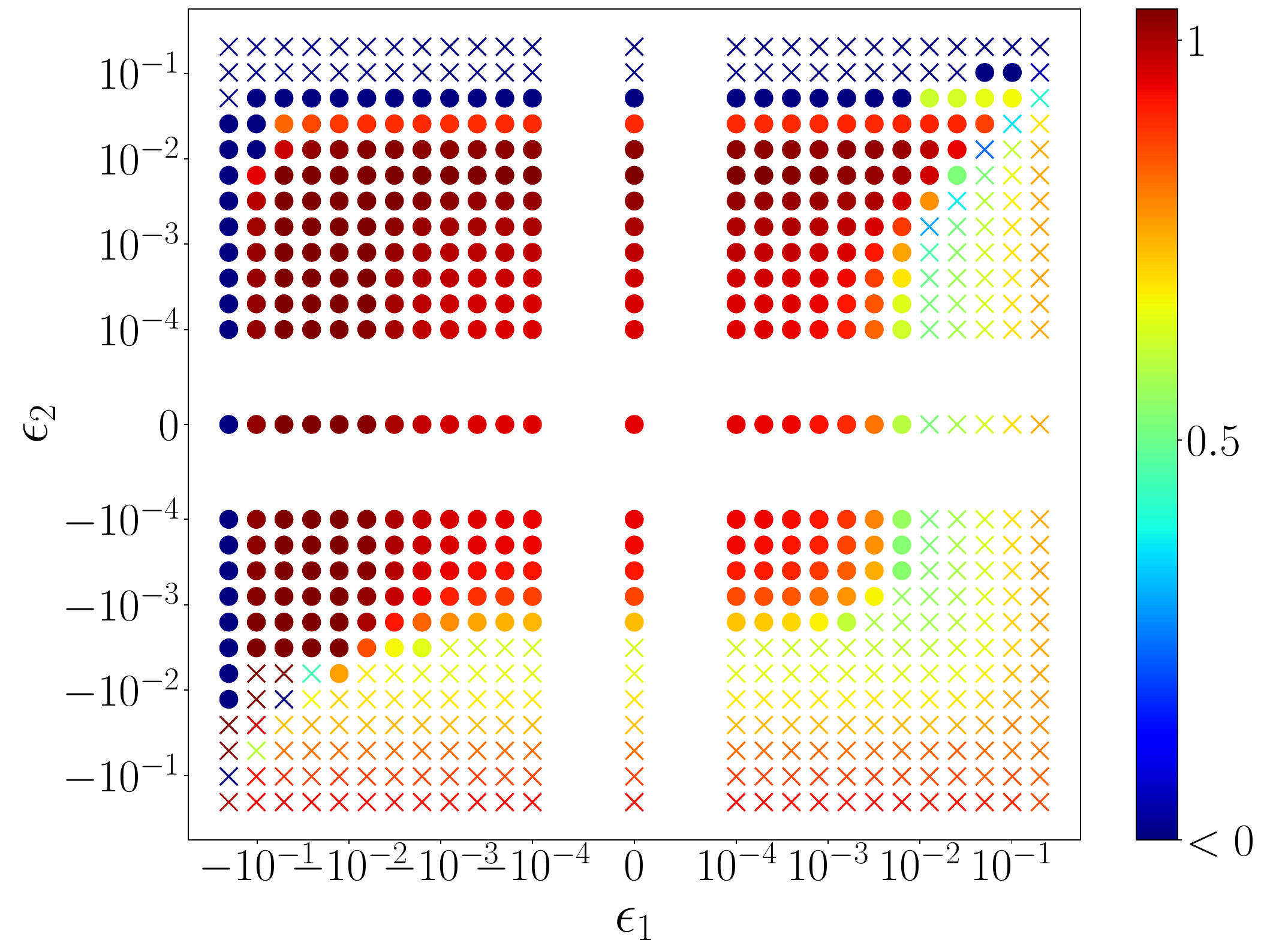}
\caption{Minimum value of $\chi_{2}$ over time and space for the collapse scenario with $A=0.0023$. Dark blue marks represent simulations where the minimum value of $\chi_{2}$ was at some point smaller than 0, making the eigenvalue complex, potentially indicating loss of well-posedness. Here crosses indicate that the simulation crashed.} 
\label{fig:chi_2}
\end{figure}

\begin{figure}[t!]
\includegraphics[width=0.45\textwidth]{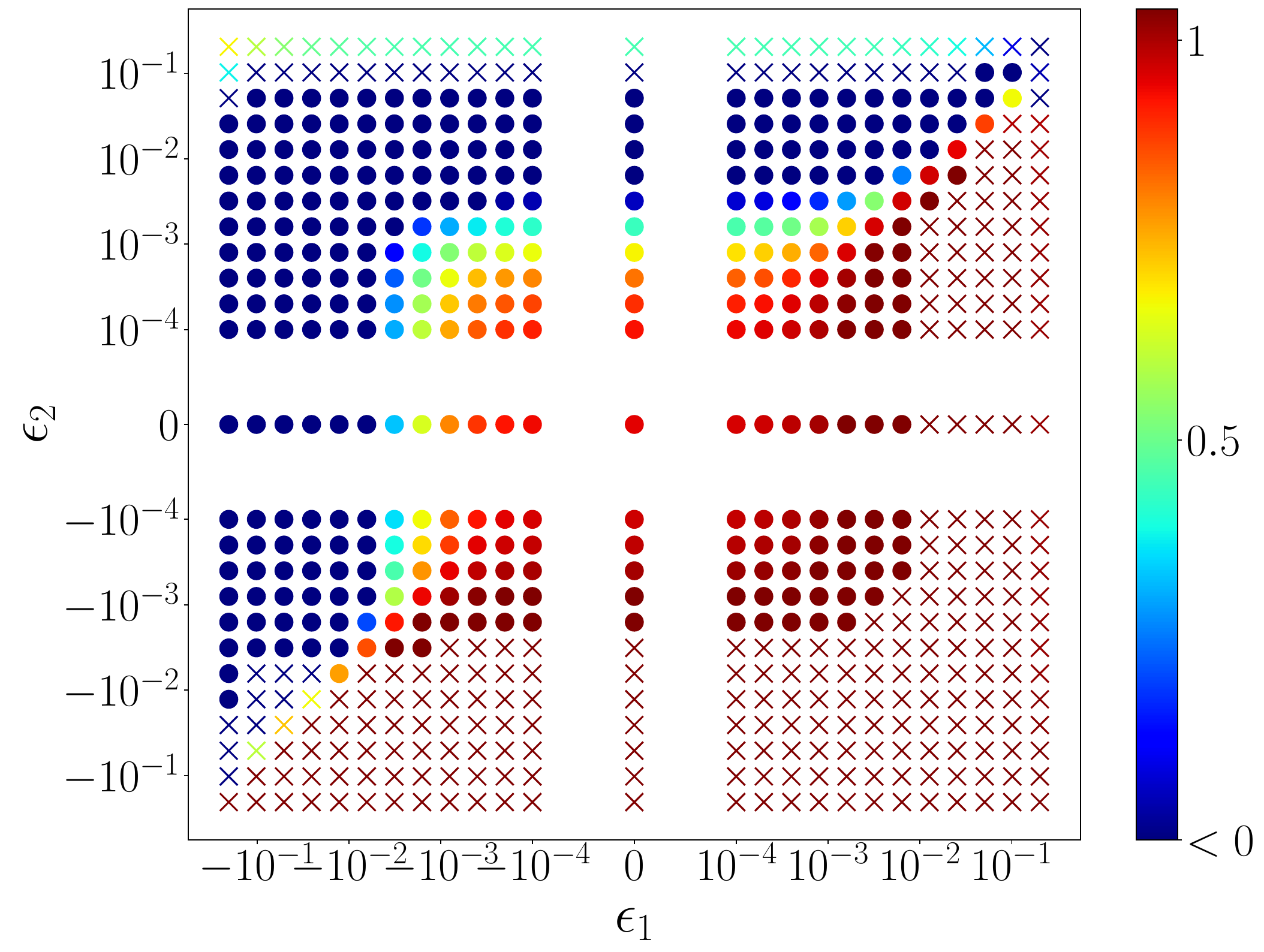}
\caption{Minimum value of $\chi_{3}$ over time and space for the collapse scenario with $A=0.0023$. Dark blue dots represent simulations where the minimum value of $\chi_{3}$ was at some point smaller than 0, making the eigenvalue complex, potentially indicating loss of well-posedness. Here crosses indicate that the simulation crashed.} 
\label{fig:chi_3}
\end{figure}

\begin{figure}[t!]
\includegraphics[width=0.45\textwidth]{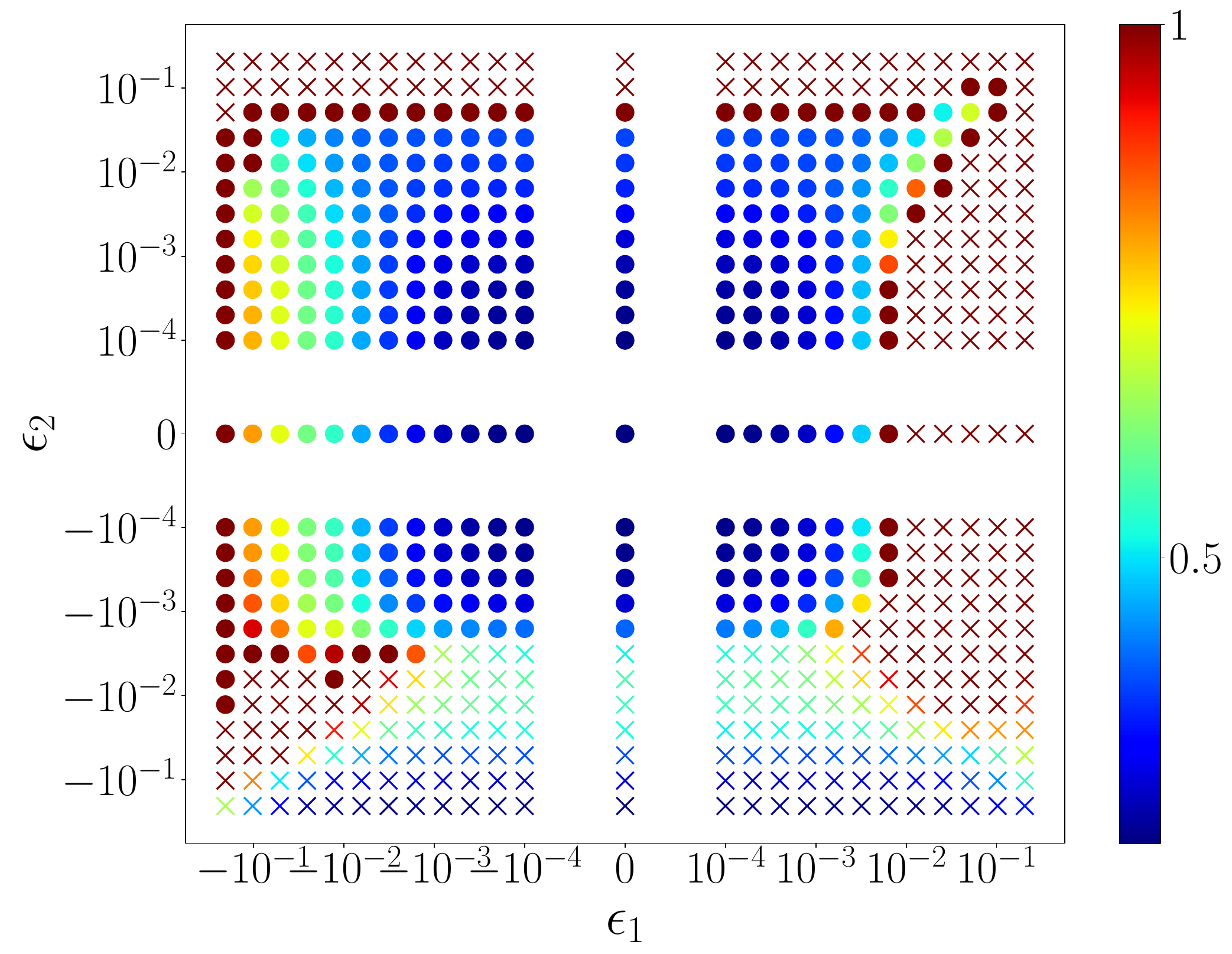}
\caption{Maximum value of $\mathcal{E}_{R}$ over time and space for the
collapse scenario with $A = 0.0023$. Dark red dots correspond to simulations where the EFT regime of applicability condition $\mathcal{E}_{R}<1$ was violated at some point. Here crosses indicate that the simulation crashed.} 
\label{fig:RICCI_EFT}
\end{figure}

\begin{figure}[t!]
\includegraphics[width=0.45\textwidth]{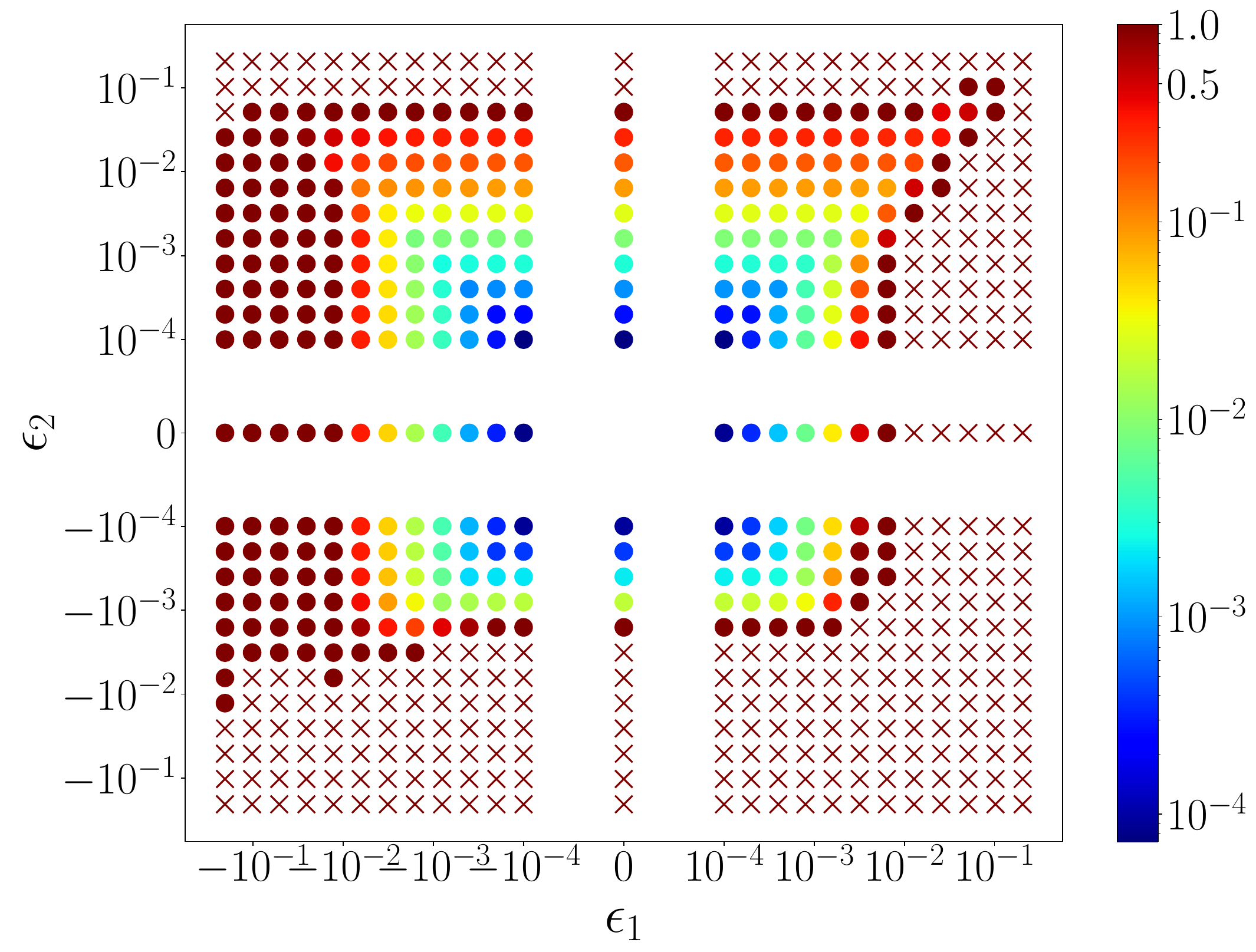}
\caption{Maximum value of $\mathcal{E}_{\mathcal{C}}$ over time and space for the
collapse scenario with $A = 0.0023$. Dark red dots correspond to simulations where the EFT regime of applicability condition $\mathcal{E}_{\mathcal{C}}<1$ was violated at some point. Here crosses indicate that the simulation crashed.} 
\label{fig:kretch_EFT}
\end{figure}

\begin{figure}[t!]
\includegraphics[width=0.45\textwidth]{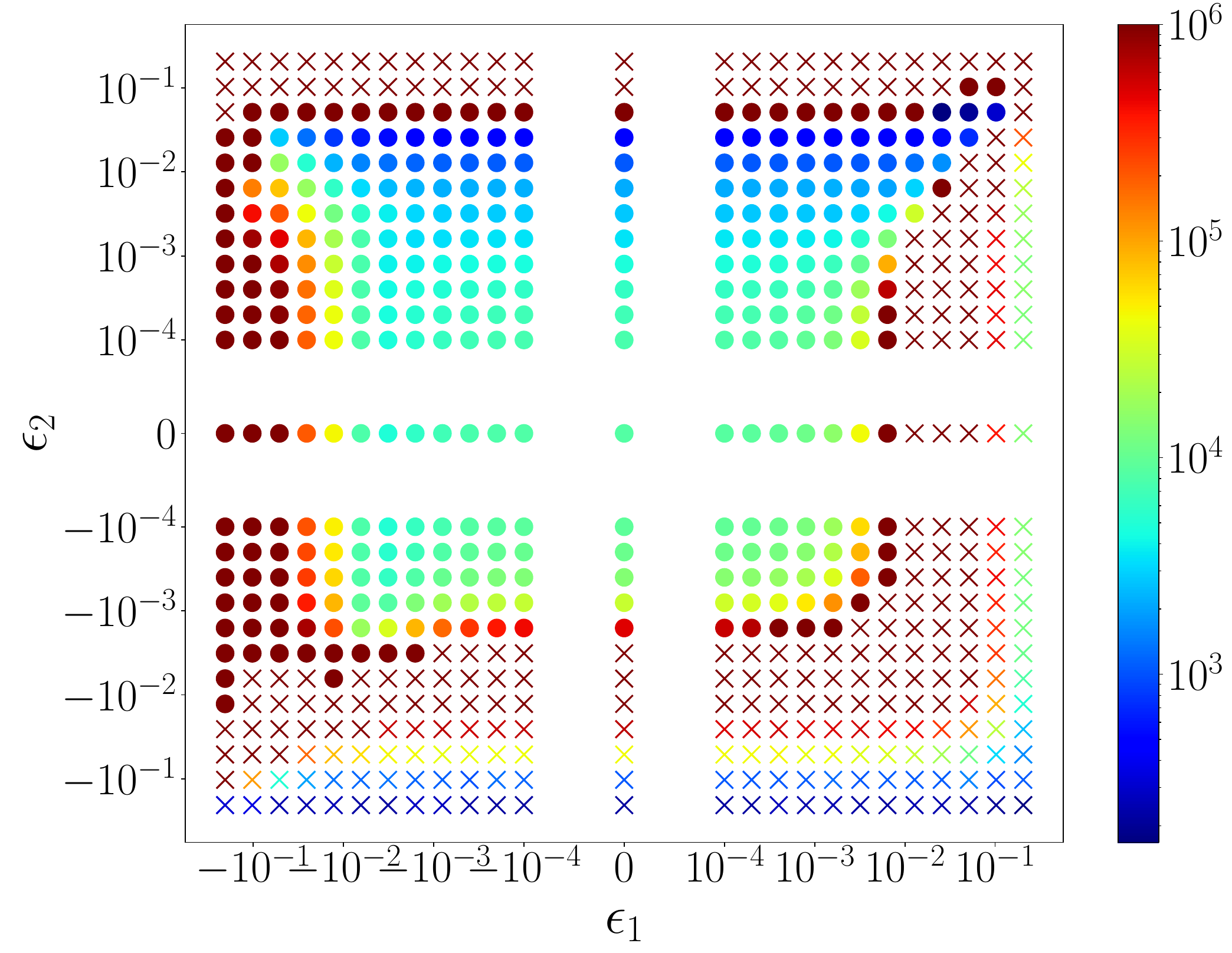}
\caption{Maximum value of $\mathcal{C}$ over time and space for the
collapse scenario with $A = 0.0023$. Values of $\mathcal{C}>10^{6}$ have been labeled as $10^{6}$ for convenience. Here crosses indicate that the simulation crashed.} 
\label{fig:max_kretch}
\end{figure}

We will not spend a lot of time going through different cases when both couplings are non-vanishing; however, informative plots are provided showing the different control quantities discussed for the $\epsilon_{1}$ and $\epsilon_{2}$ individual cases. Figure \ref{fig:chi_2} shows the space-time minimum  value of the radicand  $\chi_{2}$ of the eigenvalue $\lambda_{2\pm}$. In contrast to the previously observed for the $\chi_{3}$ quantity, when $\chi_{2}$ becomes negative, the couplings are already large enough to take the system outside the EFT regime. Figure \ref{fig:chi_3} shows the minimum space-time value of $\chi_{3}$ for each simulation. The interpretation of this plot follows directly from what was observed for the individual coupling cases. As mentioned before we can see that when $\epsilon_{1}\approx \epsilon_{2}$ simulations that would have $\chi_{3}<0$ if only $\epsilon_{2}$ was turned on, or crash if only $\epsilon_{1}$ was on, now suffer non of those issues. Similar behavior is observed for the rest of the relevant quantities.  Figure \ref{fig:RICCI_EFT} displays the maximum value of $\mathcal{E}_{R}$, on it dark red dots correspond to points where the $\mathcal{E}_{R}>1$ EFT condition was violated. Figure \ref{fig:kretch_EFT}  shows the maximum of $\mathcal{E}_{\mathcal{C}}$ over time and space; the dark red dots represent points at which the EFT condition was violated. Finally, Figure \ref{fig:max_kretch} shows the maximum space-time value of $\mathcal{C}$.

\section{Discussion}\label{discussion}

This study investigates the phenomenon of gravitational collapse in spherical symmetry within the framework of a dimension-four EFT extension to GR, commonly known as Quadratic Gravity. Within the EFT perspective, the solutions derived from this theory are expected to differ from those of GR only in the presence of matter, with the dimension-four operators representing leading-order corrections to GR within an EFT expansion.

In this particular research, instead of treating the additional degrees of freedom associated with higher derivatives as massive spin-0 and spin-2 modes, as done in previous studies such as \cite{Held:2023aap, Held:2021pht} under Ricci-flat (vacuum) scenarios, an "Order Reduction" technique \cite{Solomon:2017nlh}  is employed to eliminate these degrees of freedom. Through numerical simulations, this work is able to dynamically form BHs from the collapse of a scalar field. In addition, we identify a parameter space regime where the system is well-behaved and remains within the applicable range of the EFT. However, strong deviations in the dynamics of curvature invariants during the collapse are observed within this regime. These deviations could be particularly relevant in astrophysical scenarios like the merger of a pair of neutron stars, where the altered system dynamics could have discernible effects on the emission of gravitational radiation. The study of neutron stars for individual and binary cases in this EFT extension to GR will be explored in future work.

Additionally, instances were found where simulations, initially showing good behavior, venture into high-curvature regimes that exceed the limits of the EFT approximation. In such cases, it becomes necessary to acknowledge the inadequacy of the chosen approach in describing the system dynamics within those specific scenarios. The specific value of the couplings $\epsilon_{1}$ and $\epsilon_{2}$ (consequently the value $\Lambda$) at which this will be the case is entirely dependent on the characteristics and relevant scales in the system\footnote{ For instance, allowing the scalar pulse to have a larger width, while adjusting the amplitude to keep the ADM mass fixed, allows to carry out stable simulations that stay within the limits of the EFT for larger values of $\epsilon_{1}$ and $\epsilon_{2}$.}. Furthermore, specific regimes were identified where the system exhibits instabilities before the validity of the EFT description ceases. In these cases, alternative approaches such as "fixing the equations" may be implemented to mitigate the emergence of instabilities and control higher frequencies. This treatment will be explored in the single neutron star and neutron star binary scenarios in future work.

\section{Acknowledgements}
I thank Miguel Bezares, Pablo A. Cano, Guillaume Dideron, Pau Figueras, Aaron Held, Guillermo Lara, and Luis Lehner for valuable discussions. This work was supported in part by Perimeter Institute
for Theoretical Physics. Research at Perimeter Institute is supported by the Government of Canada through the Department of Innovation, Science and Economic
Development Canada and by the Province of Ontario through the Ministry of Economic Development, Job Creation and Trade.

\appendix

\section{Convergence}\label{app:convergence}

\begin{figure}[t!]
\includegraphics[width=0.45\textwidth]{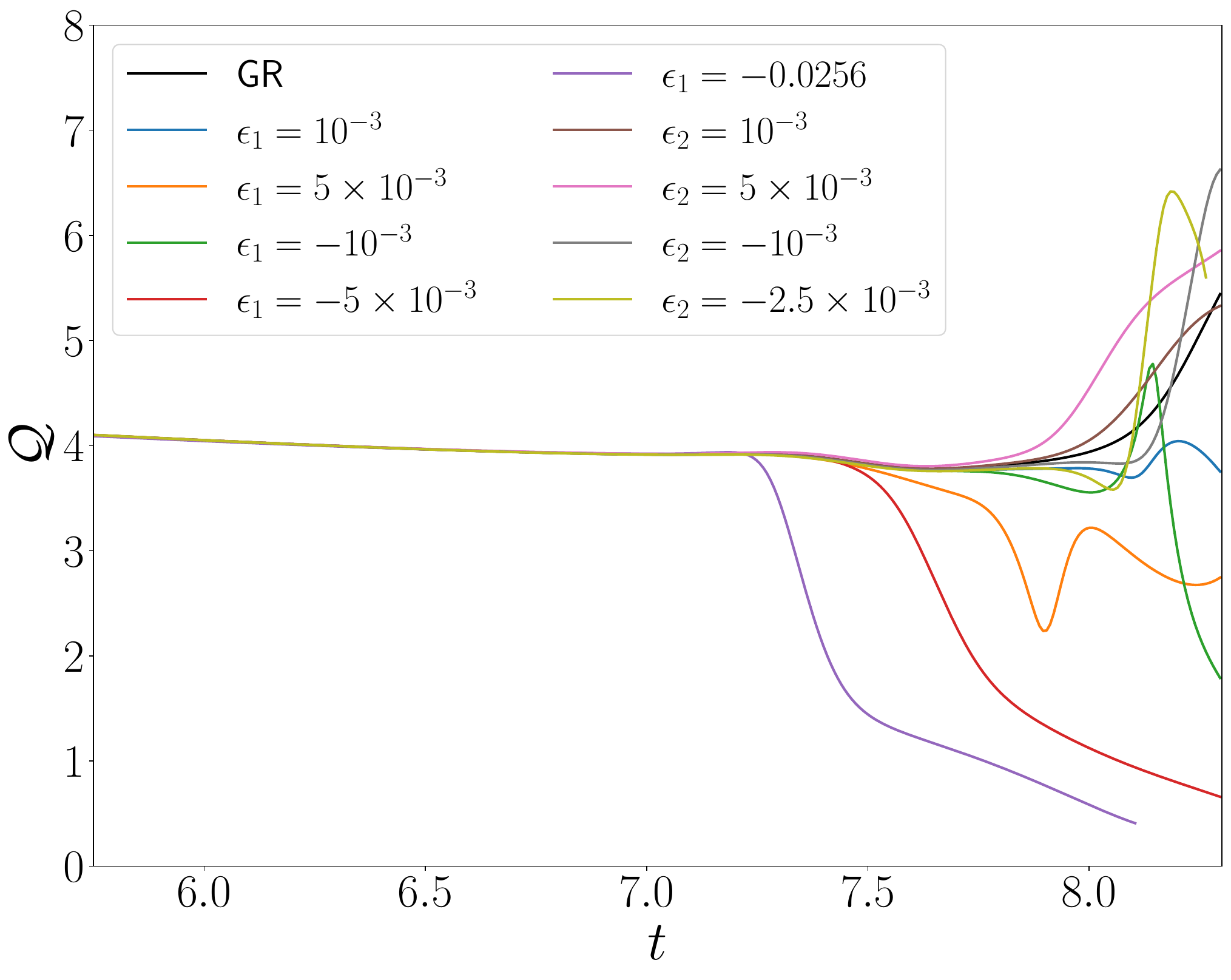}
\caption{Convergence factor $\mathcal{Q}$ for the $K_{rr}$ variable as a function of time close the time of collapse for different values of $\epsilon_{1}$ and $\epsilon_{2}$.} 
\label{fig:convergence}
\end{figure}

To check the convergence of the solutions, the base uniform grid spacing $dx=0.04$ is adopted, and the convergence factor is computed as,

\begin{equation}
    \mathcal{Q} \equiv \ln{\left( \frac{||u_{dx} - u_{dx/2} ||_{2}}{||u_{dx/2} - u_{dx/4} ||_{2}}\right)},
\end{equation}
here, $u_{dx}$, $u_{dx/2}$ and $u_{dx/4}$ stand for any field evolved with resolutions $dx$, $dx/2$ and $dx/4$ respectively. In Figure \ref{fig:convergence} we plot the convergence factor $\mathcal{Q}$ for the $K_{rr}$ variable in the BH collapse scenario : $A=0.0023$, $\sigma=1$, $r_{c}=10$, $z=0.5$, $\kappa=2$. For practical reasons, we only plot the convergence until an apparent horizon has been detected. The convergence factor behaves similarly to the other dynamical variables.
The black curve in Figure \ref{fig:convergence} shows the convergence factor for the GR case and shows how the convergence is $\approx 4$ at the beginning of the simulation and close to the collapse $\mathcal{Q}$ quickly climbs to values between 5 and 6. This is consistent with the 4th-order accuracy of the Runge-Kutta time integrator and the 6th-order accuracy of finite difference derivative operators. This seems to be similar for essentially all the $\epsilon_{2}\neq 0$ simulations. The result changes drastically for the $\epsilon_{1} \neq 0$ simulations, where we can see the convergence factor does drop to lower values as the system is close to collapse. Some of these simulations retain acceptable convergence factors, for example, the cases with $\epsilon_{1}=10^{-3}$, $\epsilon_{1}=5\times10^{-3}$ and $\epsilon_{1}=-10^{-3}$ drop to convergence factors of values $\mathcal{Q}\approx4$, $\mathcal{Q}\approx3$ and $\mathcal{Q}\approx2$ respectively. However, when the magnitude of $\epsilon_{1}$ increases, we can see how all convergence is quickly lost. This coincides mainly with the regime we have identified of simulations leaving the regime of applicability of the EFT.

\section{Constraints}\label{app:constrai}

\begin{figure}[t!]
\includegraphics[width=0.45\textwidth]{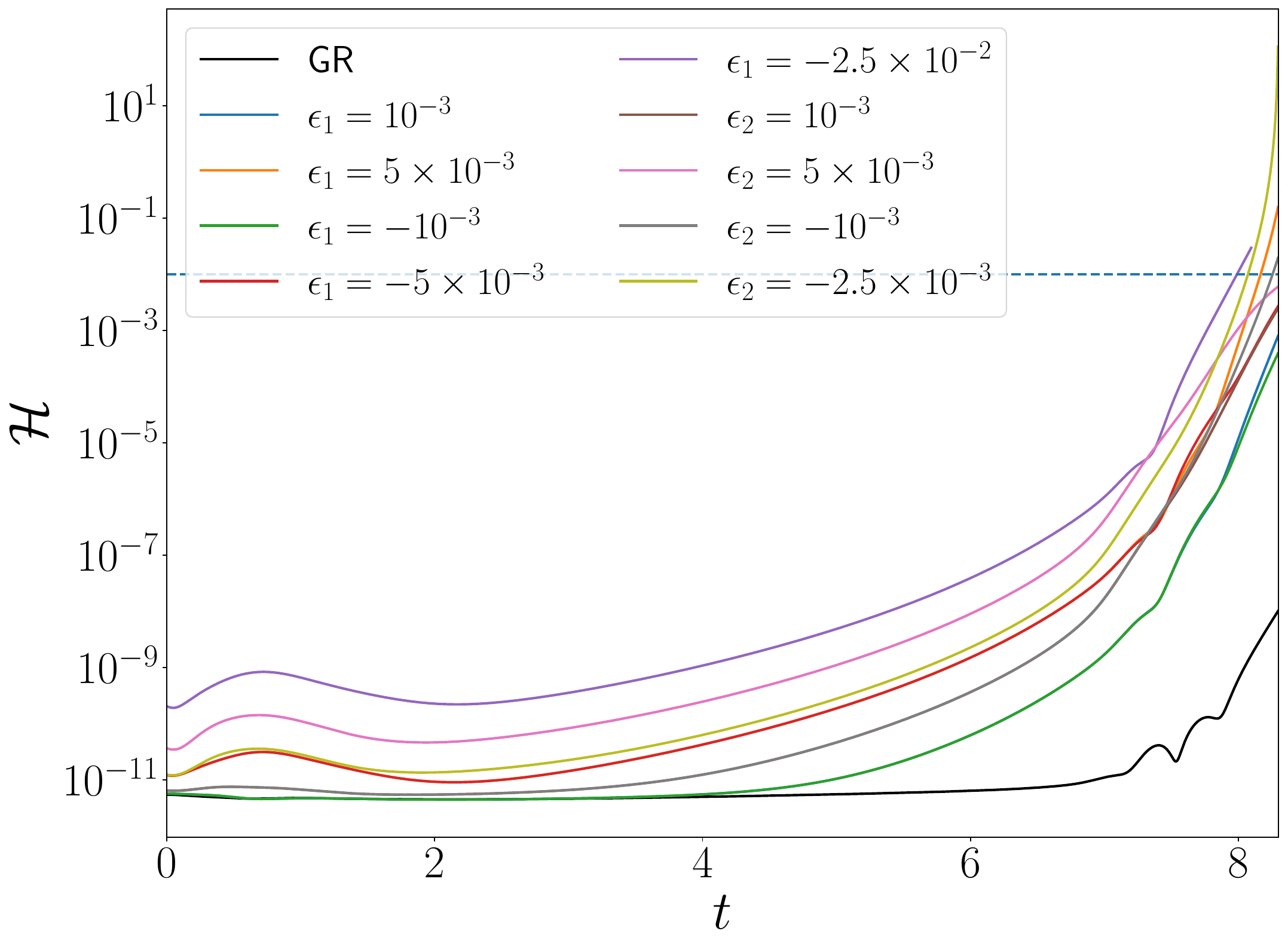}
\caption{$l2$-norm of the Hamiltonian constraint as a function of time for simulations with different values of $\epsilon_{1}$ and $\epsilon_{2}$. The horizontal dashed line highlights the $1\%$ error  .}  
\label{fig:constraints}
\end{figure}

Monitoring that the constraints \eqref{const1}, \eqref{const2}, \eqref{const3}, and \eqref{const4} remain under control is important to attest to the quality of the performed simulations. In Figure \ref{fig:constraints} we plot the valuer of the $l2$-norm of Hamiltonian constraint \eqref{const2} for different values $\epsilon_{1}$ and $\epsilon_{2}$. Here we have normalized by the $l2$-norm of the most relevant terms that define it to get a relative notion of the violation of constraints. The other constraints display similar behavior, so we omit to show them. The black curve shows our reference GR simulation using the same parameters used in the convergence test for the $dx=0.02$ grid spacing. The GR case Hamiltonian violation remains extremely small during the evolution, rising as expected close to the collapse time but never rising above a relative error of $10^{-8}$. For convenience, we only plot the constraint violations until an apparent horizon is formed; after this apparent horizon forms and excision is applied, the constraint violations naturally become smaller.

The situation changes once either of the couplings is non-vanishing; the constraint violations remain below the $10^{-8}$ relative error for most of the simulation but then quickly rise as the scalar field profile approaches the center of coordinates. For most cases, the constraint violation remains below the $1\%$ level throughout the simulation. However, there are cases in which violations are within a worrying $1\%$ and $ 10\%$ like for $\epsilon_{2}=-10^{-3}$ and $\epsilon_{1}=-2.5\times 10^{-2}$, and cases were the violations  $>10\%$  and greater than $1000\%$ error, for $\epsilon_{1}=5\times10^{3}$ and $\epsilon_{2}=-2.5\times10^{3}$. These larger constraint violations are no surprise; manipulations in the constraint equations were performed that assume that the modifying terms remain corrective (i.e., within the applicable regime of the EFT), and these corrections become greater as the pulse collapses. The cases where constraint violations are large enough to be unable to trust simulations anymore also belong in the parameter regime that has shown either through loose or convergence or by leaving the EFT regime that these solutions can not be trusted.

\newpage

\bibliographystyle{apsrev4-2}
\bibliography{main}

\end{document}